\documentclass[journal]{IEEEtran}

\usepackage{url}

\usepackage{graphicx}
\usepackage{amsmath,amssymb,amsfonts}

\usepackage[linesnumbered,boxed,ruled,commentsnumbered]{algorithm2e}

\usepackage{cite}

\usepackage{epstopdf}

\usepackage{bm}
\usepackage{booktabs}

\usepackage{subfigure}
\usepackage{enumerate}
\usepackage{makecell}

\makeatletter
\renewcommand{\maketag@@@}[1]{\hbox{\m@th\normalsize\normalfont#1}}
\makeatother

\usepackage{ragged2e}

\begin{document}

\title{Channel Acquisition for HF Skywave Massive MIMO-OFDM Communications
}

\author{Ding Shi, \IEEEmembership{Graduate Student Member, IEEE}, Linfeng Song,  \IEEEmembership{Graduate Student Member, IEEE}, \\
	Wenqi Zhou, \IEEEmembership{Student Member, IEEE}, Xiqi Gao, \IEEEmembership{Fellow, IEEE}, \\
	Cheng-Xiang Wang, \IEEEmembership{Fellow, IEEE}, and Geoffrey Ye Li, \IEEEmembership{Fellow, IEEE}
	
\thanks{
	
	This work was supported by the National Key R\&D Program of China under Grant 2018YFB1801103, the Jiangsu Province Basic Research Project under Grant BK20192002, the National Natural Science Foundation under Grant 61960206006, and the Key Research and Development Plan of Jiangsu Province (Industry Foresight and Key Core Technologies) under Grant BE2022067. Part of this work has been accepted for presentation at the IEEE GLOBECOM 2022 \cite{HFmimo}. 
	
	Ding Shi, Linfeng Song, Wenqi Zhou, Xiqi Gao and Cheng-Xiang Wang are with the National Mobile Communications Research Laboratory, Southeast University, Nanjing 210096, China, and also with Purple Mountain Laboratories, Nanjing 211111, China (e-mail: shiding@seu.edu.cn; songlf@seu.edu.cn; wqzhou@seu.edu.cn; xqgao@seu.edu.cn; chxwang@seu.edu.cn).

	Geoffrey Ye Li is with the Department of Electrical and Electronic
	Engineering, Imperial College London, London SW7 2AZ, U.K. (e-mail:
	geoffrey.li@imperial.ac.uk).
}

}

\maketitle

\begin{abstract}
In this paper, we investigate channel acquisition for high frequency (HF) skywave massive multiple-input multiple-output (MIMO) communications with orthogonal frequency division multiplexing (OFDM) modulation. 
We first introduce the concept of triple beams (TBs) in the  space-frequency-time (SFT) domain and establish a TB based channel model using sampled triple steering vectors.
With the established channel model, we then investigate the optimal channel estimation and pilot design for pilot segments.  Specifically, we find the conditions that allow pilot reuse among multiple user terminals (UTs), which significantly reduces pilot overhead and increases the number of UTs that can be served.
Moreover, we propose a channel prediction method for data segments based on the estimated TB domain channel.
To reduce the complexity, we formulate the channel estimation as a statistical inference problem and then obtain the channel by the proposed constrained Bethe free energy minimization (CBFEM) based channel estimation algorithm, which can be implemented with low complexity by exploiting the structure of the TB matrix together with the chirp z-transform (CZT).
Simulation results demonstrate the superior performance of the proposed channel acquisition approach.
\end{abstract}


\IEEEpeerreviewmaketitle

\section{Introduction}
High frequency (HF) communications, whose frequency range is usually from 3 MHz to 30 MHz, can provide worldwide coverage through skywave propagation. Compared with satellite communications, an alternative for global coverage, HF communications can be flexibly deployed with relatively low cost and are robust to jamming \cite{8456447,6316776}. However, the low data transmission rate of HF communications makes it hard to compete with satellite communications. Therefore, there has been some research on applying the  multiple-input multiple-output (MIMO) technique to traditional point-to-point HF communications \cite{7775853,7604101,6290228}, where certain performance gains can be achieved.

In the past decade, massive MIMO has been widely studied for terrestrial cellular communications. 
It can significantly improve the spectrum and power efficiencies by deploying a large number of antennas at the base station
(BS) and simultaneously serving a number of user terminals (UTs) in the same time-frequency resource \cite{6798744}. 
Recently, it has been introduced into the HF skywave communications in  \cite{yxl}, where a spatial-beam based wideband model for HF skywave massive MIMO
channels within the orthogonal frequency division multiplexing (OFDM)
transmission framework has been developed and the asymptotic
achievable sum-rate is obtained when using the minimum mean-squared
error (MMSE) based uplink (UL) receiver and downlink (DL)
precoder with perfect channel state information (CSI) at the BS.
It is demonstrated in \cite{yxl} that massive MIMO can also vastly improve the performance of HF skywave communications.

The acquisition of the CSI is essential for massive MIMO to harvest the significant performance gain, which has been well investigated in terrestrial cellular massive MIMO systems \cite{6415397,7045498,9042356,7524027,8334183,7124505,7355354,6816089,6951471,6940305,8425578,8171203,8723310,9351786,9439804,8328018,7174558,9399122,9556593,9133156,7961152}. 
The conventional orthogonal pilot signaling and channel estimation methods, such as least squares (LS) and MMSE \cite{6415397}, are not suitable for the massive MIMO system due to
the overwhelming pilot overhead and the forbidding computational complexity.
By exploiting the constraint of the angle spread in the massive MIMO, the pilot reuse schemes have been proposed to reduce the pilot overhead, where UTs with non-overlapping angle domain channels are allowed to share the same pilot \cite{7045498,9042356,7524027,8334183}.  
For these pilot schemes, the pilot scheduling can be performed based on channel statistics \cite{7045498,9042356} or  the spatial information acquired by the spatial rotation enhanced discrete Fourier transform (DFT) method \cite{7524027,8334183}. 
On the other hand, due to the limited local scatters, the terrestrial massive MIMO channel exhibits sparsity \cite{zhou2007experimental}, which enables channel estimation through sparse signal recovery. Thus sparse channel estimation in \cite{5454399} has been proposed to reduce the pilot overhead. 
In \cite{7124505,7355354,6816089,6951471}, the greedy algorithms  iteratively identify the sparse support and reconstruct the channel. 
Besides, statistical inference methods have also been developed for efficient sparse channel recovery.
The expectation-maximization (EM) Gaussian-mixture approximate message passing (AMP) algorithm \cite{6940305} and the turbo-orthogonal AMP algorithm \cite{8425578} exploit the channel sparsity in the angle domain to estimate massive MIMO channels. For massive MIMO-OFDM communications, the sparse angle-delay domain channel can be obtained by EM-generalized AMP, EM-vector AMP algorithms \cite{8171203} and the structured turbo compressive sensing (CS) algorithm \cite{8723310}.
Recently, constrained Bethe free energy minimization (CBFEM) unifies different message passing algorithms into a single optimization framework in \cite{9351786}. CBFEM transforms the statistical inference problem into an optimization one with a clear objective function, where different constraints will lead to different solving algorithms, such as the expectation propagation (EP) variant \cite{9351786}, the hierarchical hybrid message passing \cite{9439804}, and the AMP with nearest neighbor sparsity pattern learning algorithm \cite{8328018}.
Moreover, channel estimation approaches  highly depend on the channel models. The spatial-beam based channel model has been widely used in \cite{7045498,6940305,7174558} for channel estimation and also applied to the signal detection \cite{7794626} and precoding \cite{8694866}. An accurate spatial-beam based channel model using finely sampled steering vectors in \cite{9399122,9556593,9133156,7961152} can improve channel estimation performance.

In this paper, we investigate channel acquisition for HF skywave massive MIMO-OFDM communications. We introduce the concept of the triple beams (TBs) in space-frequency-time (SFT) domain and use it to derive a TB based channel model. Based on the TB based channel model, we can design pilots and develop the channel estimation algorithm to accurately estimate the HF skywave massive MIMO-OFDM channels.
The main contributions are summarized as follows.

\begin{itemize}
	\item We derive a TB based channel model from the physical principle of HF skywave channels using sampled triple steering vectors in the SFT domain, each of which corresponds to a physical TB consisting of the spatial-beam, the frequency-beam, and the temporal-beam, pointing towards the sampled directional cosine, delay, and Doppler frequency, respectively. 
	
	\item Based on the proposed channel model, we first investigate the optimal estimation of channels at pilot segments and find the conditions to minimize the normalized mean-squared error (NMSE) of channel estimates. Specifically, we show that UTs with overlapping TB domain channels should be allocated pilot sequences with different phase shift factors while UTs with non-overlapping TB domain channels can reuse the same pilot sequence. Furthermore, the pilot is designed, including UT grouping and pilot scheduling, and the channel prediction method for data segments is established based on the estimated TB domain channel.

	\item We formulate the channel estimation as a statistical inference problem, which can be solved within the optimization framework of CBFEM. We then acquire the channel by the proposed CBFEM based channel estimation algorithm, which can be implemented with low complexity by exploiting the structure of the TB matrix together with the chirp z-transform (CZT).

\end{itemize}

The rest of the paper is organized as follows. 
In Section II, we derive the channel model for HF skywave massive MIMO-OFDM communications.
In Section III, we investigate the optimal channel estimation for pilot segments and develop a pilot design. Moreover, a channel prediction method for data segments is presented.
In Section IV, the CBFEM based channel estimation algorithm and its low-complexity implementation are presented.
Section V provides simulation results and the paper is concluded in Section VI.

\emph{Notations}: 
The uppercase (lowercase) boldface letters denote matrices (vectors).
The superscripts ${\left(  \cdot  \right)^ * }$, ${\left(  \cdot  \right)^ {\rm{T}} }$, ${\left(  \cdot  \right)^{\rm{H}}}$ and ${\rm{tr}}\left\{ {\cdot} \right\}$ denote the conjugate, transpose, conjugated-transpose, and matrix trace operations, respectively.
${\bar \jmath} = \sqrt { - 1} $ denotes the imaginary unit.
${\left\| {\bf{x}} \right\|}$ denotes the $\ell_2$-norm of ${\bf{x}}$, and ${\left\| {\bf{X}} \right\|_{\rm{F}}}$ denotes the Frobenius norm of ${\bf{X}}$.
$\left| \Psi  \right|$ denotes the cardinality of set $\Psi$. 
${\Psi _1} \times {\Psi _2}$ denotes the Cartesian product of sets ${\Psi _1}$ and ${\Psi _2}$.
$\lceil x \rceil$ denotes the smallest integer that is not less than $x$, while $\left\lfloor x \right\rfloor $ denotes the largest integer that is not greater than $x$.
$ \circ $ and $ \otimes $ denote the Hadamard and the Kronecker product operators, respectively.
${\left[ {\bf{x}} \right]_a}$ and ${\left[ {\bf{x}} \right]_{a:b}}$ denote the $a$-th element and elements $a$ to $b$ of vector ${\bf{x}}$, respectively.
${\left[ {\bf{X}} \right]_{a,b}}$, ${\left[ {\bf{X}} \right]_{:,b}}$, and ${\left[ {\bf{X}} \right]_{a:b,c:d}}$ denote the $(a,b)$-th element, the $b$-th colomn, and rows $a$ to $b$ and columns $c$ to $d$ of ${\bf{X}}$, respectively.
$\mathbb{E}\left\{  \cdot  \right\}$, $\text{Var}\left\{  \cdot  \right\}$, $\mathbb{H}\left\{  \cdot  \right\}$ and $\mathbb{D}\left\{  \cdot  \right\}$ denote the statistical
expectation, variance, entropy, and relative entropy, respectively.
${\rm{diag}}\left\{ {\bf{x}} \right\}$ denotes the diagonal matrix with $\bf{x}$ along its main diagonal, and ${\rm{diag}}\left\{ {{{\bf{X}}_1}, \cdots ,{{\bf{X}}_N}} \right\}$ forms the block-diagonal matrix with ${{{\bf{X}}_1}, \cdots ,{{\bf{X}}_N}}$.
${\bf{0}}$ denote the all-zero vector or matrix. 
${{\bf{I}}_N}$ denotes the identity matrix of dimension $N$ while ${{\bf{I}}_{N \times G}}$ denotes the matrix composed of the first $G$ ($ \le N$) columns of ${{\bf{I}}_N}$.
${\bf{X}} \succ {\bf{0}}$ (${\bf{X}} \succeq {\bf{0}}$) denotes that $\bf{X}$ is Hermitian positive definite (semi-definite). 
$\mathcal{CN}\left( {{\bf{x}};{\bf{a}},{\bf{A}}} \right)$ denotes the
circular symmetric complex Gaussian probability density function with mean
$\bf{a}$ and covariance $\bf{A}$.
\section{HF Skywave massive MIMO-OFDM Channel Model}

After introducing the configuration of the HF skywave massive MIMO-OFDM system in this section, we present the concept of the triple beams in the SFT domain and establish a triple-beam based channel model from the physical principle of HF skywave channels by using sampled triple steering vectors.

\subsection{System Configuration}
We consider a HF skywave massive MIMO-OFDM system. The BS is equipped with a uniform linear array (ULA) with $M$ antennas and serves $U$ single-antenna UTs. Let $N_{\rm{c}}$ denote the number of subcarriers, $N_{\rm{g}}$ denotes the length of the cyclic prefix (CP), $\Delta f$ denotes the subcarrier spacing, and ${T_{\rm{s}}} = \frac{1}{{{N_{\rm{c}}}\Delta f}}$ denotes the sampling interval. We assume that $N_{\rm{v}}$ valid subcarriers are used to transmit data and pilots with the index set $\mathcal{K}=\left\{ {{k_0},{k_1}, \cdots ,{k_{{N_{\rm{v}}} - 1}}} \right\}$.

The HF skywave massive MIMO-OFDM system operates in time-division duplex (TDD) mode with frame structure shown in Fig. 1. 
Each frame consists of ${N_{\rm{F}}}$ timeslots and each timeslot consists of ${N_{\rm{S}}}$ OFDM symbols, thus the number of OFDM symbols in each frame is $N = {N_{\rm{F}}}{N_{\rm{S}}}$. In each timeslot, the $n_{\rm{p}}$-th OFDM symbol is the pilot used for UL training and the other OFDM symbols are used for UL/DL data transmission.

\begin{figure}[htbp]
	\centering
	\includegraphics[width=0.50\textwidth]{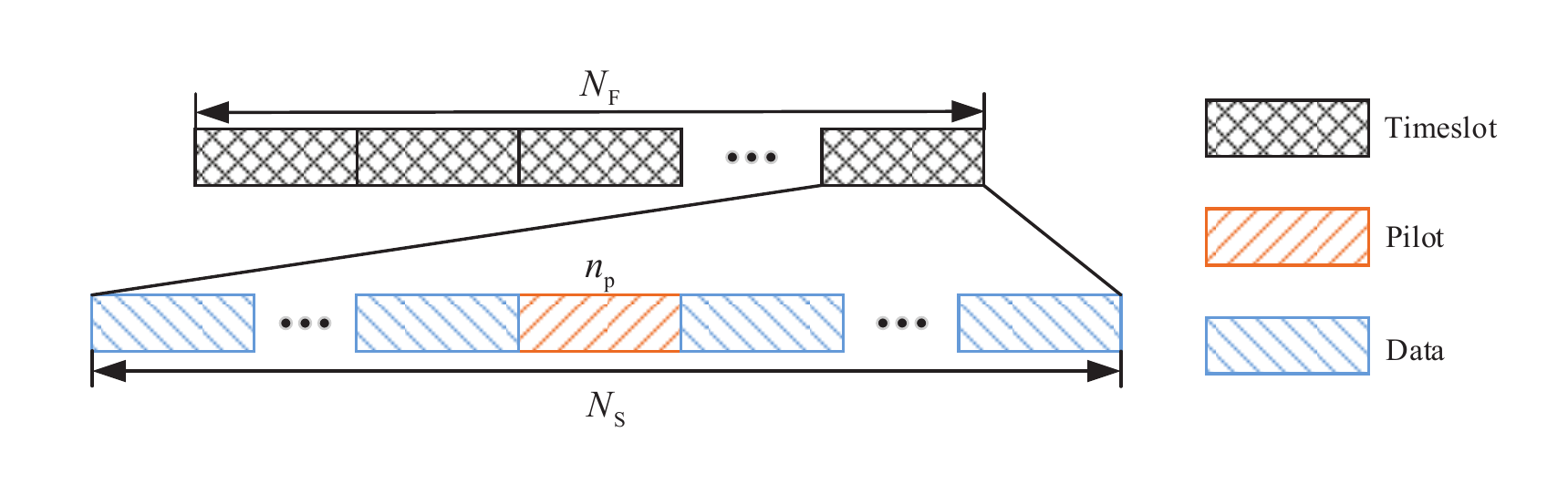}
	\caption{The frame structure.}
	\label{fig-TDD}
\end{figure}

In HF skywave communications, the carrier frequency $f_{\rm{c}}$ varies with the ionospheric conditions at different times. 
Therefore, we set inter-antenna spacing $d$ according to the highest system operating frequency $f_{\rm{o}}$, that is,
$d = {{{\lambda _{\rm{o}}}} \mathord{\left/{\vphantom {{{\lambda _{\rm{o}}}} 2}} \right. \kern-\nulldelimiterspace} 2}$, where ${\lambda _{\rm{o}}} = {c \mathord{\left/	{\vphantom {c {{f_{\rm{o}}}}}} \right.	\kern-\nulldelimiterspace} {{f_{\rm{o}}}}}$ is the wavelength, and $c$ is the speed of light. This is different from the traditional massive MIMO communications, where the antenna spacing is usually half the wavelength of the carrier.

\subsection{Triple-Beam Based Channel Model}
Let ${x_{u,n,k}}$ denote the transmitted data of UT $u$ on the $n$-th OFDM symbol at the $k$-th subcarrier, where $n \in \left\{ {0,1, \cdots ,N - 1} \right\}$ and $k \in \mathcal{K}$. After OFDM modulation, the analog baseband signal on the $n$-th OFDM symbol transmitted by UT $u$ with the CP can be expressed as
\begin{equation}
{\bar x_{u,n}}\!\left( t \right) \!= \!\!\!\! \sum\limits_{k = {k_0}}^{{k_{{N_{\rm{v}}} - 1}}} \!\!\!{x_{u,n,k}{e^{  {\bar \jmath}2\pi k\Delta ft}}} , - {N_{\rm{g}}}{T_{\rm{s}}} \!\le\! t \!- n{T_{{\rm{sym}}}}\! <\! {N_{\rm{c}}}{T_{\rm{s}}},
\end{equation}
where ${T_{{\rm{sym}}}} = \left( {{N_{\rm{c}}} + {N_{\rm{g}}}} \right){T_{\rm{s}}}$ is the time duration of an OFDM symbol including CP. At the BS, the received analog baseband signal on the $n$-th OFDM symbol at the $m$-th antenna can be expressed as
\begin{equation}
{\bar y_{m,n}}\left( t \right) \!= \!\sum\limits_{u = 0}^{U - 1}\! {\int_{ - \infty }^\infty \!\! {{{\bar h}_{u,m}}\left( {t,\tau } \right){{\bar x}_{u,n}}\left( {t - \tau } \right)d\tau } }  + {\bar z_{m,n}}\left( t \right),
\end{equation}
where $m \in \left\{ {0,1, \cdots ,M - 1} \right\}$, 
${\bar z_{m,n}}\left( t \right)$ is the additive white Gaussian noise and ${\bar h_{u,m}}\left( {t,\tau } \right)$ is the time-varying channel impulse response between  UT $u$ and the $m$-th antenna of the BS. The channel impulse response can be expressed as
\begin{align}\label{equ-chan}
{\bar h_{u,m}}\left( {t,\tau } \right) &= \sum\limits_{p = 0}^{{P_u} - 1} {\gamma _{u,p}}{e^{{\bar \jmath}2\pi {\nu _{u,p}}t}}{e^{ - {\bar \jmath}2\pi {f_{\rm{c}}}m\Delta \tau {\Omega _{u,p}}}} \nonumber \\
&\qquad\qquad\times \delta \left( {\tau  - {\tau _{u,p}} - m\Delta \tau {\Omega _{u,p}}} \right) ,
\end{align}
where $P_u$ is the number of paths between UT $u$ and the BS, ${\gamma _{u,p}}$, ${{\nu_{u,p}}}$, and ${\Omega _{u,p}}$ is the complex-valued gain, the Doppler frequency, and the directional cosine of the $p$-th path of UT $u$, respectively, ${\tau _{u,p}}$ is the $p$-th propagation path delay between UT $u$ and the first antenna of the BS, and $\Delta \tau  = {d \mathord{\left/	{\vphantom {d c}} \right.	\kern-\nulldelimiterspace} c}$. In (\ref{equ-chan}), the complex-valued gain can be expressed as ${\gamma _{u,p}} = {\beta _{u,p}}{e^{{\bar \jmath}{\varphi _{u,p}}}}$, where ${\beta _{u,p}}$ and ${{\varphi _{u,p}}}$ are the gain and initial phase, respectively, and ${{\varphi _{u,p}}}$ is uniformly distributed over $\left[ {0,2\pi } \right)$. The directional cosine is defined as ${\Omega _{u,p}}  \buildrel \Delta \over =  \sin \theta _{u,p}^{{\rm{az}}}\cos \theta _{u,p}^{{\rm{el}}}$, where $\theta _{u,p}^{{\rm{az}}}$ and $\theta _{u,p}^{{\rm{el}}}$ are the azimuth angle of arrival (AoA) and elevation AoA, respectively.
Note that the propagation delay across the antenna array (i.e., ${m\Delta \tau {\Omega _{u,p}}}$ in the delta function) is considered due to the spatial-wideband effect caused by the equipped large-scale antenna array and wider bandwidth compared with traditional HF communications \cite{8354789,yxl}.

We assume that the channel state keeps constant within an OFDM symbol and varies symbol by symbol due to the Doppler effect. After the OFDM demodulation, the received data at the $k$-th subcarrier on the $n$-th OFDM symbol at the $m$-th antenna is given by
\begin{equation}\label{equ-signal1}
{y_{m,n,k}} = \sum\limits_{u = 0}^{U - 1} {h_{u,m,n,k}^{{\rm{SFT}}}x_{u,n,k}}  + {z_{m,n,k}},
\end{equation}
where ${z_{m,n,k}}$ is the additive white Gaussian noise with distribution $\mathcal{CN}\left( {{z_{m,n,k}};0,\sigma _{\rm{z}}^2} \right)$, and ${h_{u,m,n,k}^{{\rm{SFT}}}}$ 
is channel frequency response at the $k$-th subcarrier on the $n$-th OFDM symbol between UT $u$ and the $m$-th antenna at the BS, which is given by
\begin{align}\label{equ-chan1}
h_{u,m,n,k}^{{\rm{SFT}}}&= \int {{{\bar h}_{u,m}}\left( {n{T_{{\rm{sym}}}},\tau } \right)} {e^{ - {\bar \jmath}2\pi k\Delta f\tau }}d\tau \nonumber\\
&= \sum\limits_{p = 0}^{{P_u} - 1} {\gamma _{u,p}}{e^{{\bar \jmath}2\pi {\nu _{u,p}}n{T_{{\rm{sym}}}}}}{e^{ - {\bar \jmath}2\pi {f_{\rm{c}}}m\Delta \tau {\Omega _{u,p}}}} \nonumber \\
& \qquad\qquad\times{e^{ - {\bar \jmath}2\pi k\Delta f{\tau _{u,p}}}}{e^{ - {\bar \jmath}2\pi k\Delta fm\Delta \tau {\Omega _{u,p}}}} .
\end{align}

We consider the whole space-frequency domain channel over $N$ OFDM symbols between UT $u$ and the BS, which is referred to as the space-frequence-time (SFT) domain channel vector, ${\bf{h}}_u^{{\rm{SFT}}}$, with element $h_{u,m,n,k}^{{\rm{SFT}}}$ of index $\left( {nM{N_{\rm{v}}} + \left( {k - {k_0}} \right)M + m} \right)$. We denote
\begin{align}
{{\bf{v}}_k}\left( {{\Omega}} \right) \buildrel \Delta \over = &\Big[ 1,{e^{ - {\bar \jmath}2\pi \left( {{f_{\rm{c}}} + k\Delta f} \right)\Delta \tau {\Omega}}}, \cdots , \nonumber \\
&\qquad\qquad{e^{ - {\bar \jmath}2\pi \left( {{f_{\rm{c}}} + k\Delta f} \right)\left( {M - 1} \right)\Delta \tau {\Omega}}} \Big]^{\rm{T}}\!\!\! \in {\mathbb{C}^{M \times 1}},
\end{align}
\begin{align}
{\bf{u}}\left( {{\tau}} \right) \buildrel \Delta \over = &\Big[ {e^{ - {\bar \jmath}2\pi {k_0}\Delta f{\tau}}},{e^{ - {\bar \jmath}2\pi {k_1}\Delta f{\tau}}}, \cdots , \nonumber \\
&\qquad\qquad\qquad\qquad{e^{ - {\bar \jmath}2\pi {k_{{N_{\rm{v}}} - 1}}\Delta f{\tau}}} \Big]^{\rm{T}} \!\!\in {\mathbb{C}^{{N_{\rm{v}}} \times 1}},
\end{align}
\begin{equation}
{\bf{d}}\left( {{\nu }} \right) \buildrel \Delta \over = {\left[ {1,{e^{{\bar \jmath}2\pi {\nu }{T_{{\rm{sym}}}}}}, \cdots ,{e^{{\bar \jmath}2\pi {\nu }\left( {N - 1} \right){T_{{\rm{sym}}}}}}} \right]^{\rm{T}}} \in {\mathbb{C}^{N \times 1}},
\end{equation}
as steering vectors in the space domain, the frequency domain, and the time domain,  pointing towards directional cosine ${{\Omega }}$, delay ${{\tau }}$, and Doppler frequency ${{\nu }}$, respectively. Note that the steering vectors in the space domain are different for different subcarriers due to the spatial-wideband effect. Denote
\begin{equation}\label{equ-tb-vector}
{\bf{p}}\left( {{\Omega },{\tau },{\nu }} \right)\! \buildrel \Delta \over =\! {\bf{d}}\left( {{\nu }} \right) \otimes \left( {{\bf{v}}\left( {{\Omega }} \right)\! \circ \!\left( {{\bf{u}}\left( {{\tau }} \right)\! \otimes\! {{\bf{e}}_M}} \right)} \right) \!\in\! {\mathbb{C}^{M{N_{\rm{v}}}N\! \times 1}},
\end{equation}
where ${{\bf{e}}_M} \buildrel \Delta \over = {\left[ {1,1, \cdots ,1} \right]^{\rm{T}}} \in {\mathbb{C}^{M \times 1}}$, and 
\begin{equation}
{\bf{ v}}\left( {{\Omega }} \right) \!\buildrel \Delta \over =\! {\left[ {{{\bf{v}}_{{k_0}}}{{\left( {{\Omega }} \right)}^{\!\rm{T}}}\!,{{\bf{v}}_{{k_1}}}{{\left( {{\Omega }} \right)}^{\!\rm{T}}}\!, \!\cdots\! ,{{\bf{v}}_{{k_{{N_{\rm{v}}}\! - \!1}}}}{{\left( {{\Omega }} \right)}^{\!\rm{T}}}} \right]^{\!\rm{T}}} \!\!\!\!\in\! {\mathbb{C}^{M{N_{\rm{v}}} \!\times\! 1}}.
\end{equation}
Therefore, the SFT domain channel can be expressed as
\begin{equation}\label{equ-chan2}
{\bf{h}}_u^{{\rm{SFT}}} = \sum\limits_{p = 0}^{{P_u} - 1} {{\gamma _{u,p}}{\bf{p}}\left( {{\Omega _{u,p}},{\tau _{u,p}},{\nu _{u,p}}} \right)}  \in {\mathbb{C}^{M{N_{\rm{v}}}N \times 1}}.
\end{equation}
In this physical channel model, ${\bf{p}}\left( {{\Omega _{u,p}},{\tau _{u,p}},{\nu _{u,p}}} \right)$ represents a triple steering vector pointing towards the channel parameters $\left( {{\Omega _{u,p}},{\tau _{u,p}},{\nu _{u,p}}} \right)$. Moreover, the complex-valued gain of a path, ${{\gamma _{u,p}}}$, is also the weight of the triple steering vector.
Based on this physical channel model, we next derive a statistical channel model by sampling the triple steering vectors, which is referred to as the triple-beam based channel model and will be used in the channel acquisition. 

We assume that all UTs are synchronized. Note that the parameters ${\left( {{\Omega _{u,p}},{\tau _{u,p}},{\nu _{u,p}}} \right)}$ of each path for each UT are limited within sets ${\mathcal{B}_{{\rm{an}}}} = \left\{ {\Omega \left| {\Omega  \in \left[ { - 1,1} \right)} \right.} \right\}$, ${\mathcal{B}_{{\rm{de}}}} = \left\{ {\tau \left| {\tau  \in \left[ {0,{\tau _{\max }}} \right)} \right.} \right\}$ and ${\mathcal{B}_{{\rm{do}}}} = \left\{ {\nu \left| {\nu  \in \left[ { - {\nu _{\max }},{\nu _{\max }}} \right)} \right.} \right\}$, respectively, where ${\tau _{\max }}  \buildrel \Delta \over =  \frac{{{N_{\rm{g}}}}}{{{N_{\rm{c}}}\Delta f}}$ is the maximum delay spread, and ${\nu _{\max }}  \buildrel \Delta \over =  \frac{{{N_{\rm{d}}}}}{{2N{T_{{\rm{sym}}}}}}$ is the maximum Doppler \cite{8727425,9440710}. We uniformly divide these sets into multiple disjoint subsets as
\begin{subequations}\label{equ-sam_int}
\begin{equation}
{\mathcal{B}_{{\rm{an}}}}\!\! =\!\!\! \bigcup\limits_{{n_{{\rm{an}}}} = 0}^{{N_{{\rm{an}}}} - 1}\!\!\!\! {\Lambda _{{n_{{\rm{an}}}}}^{{\rm{an}}}} ,\   \Lambda _{{n_{{\rm{an}}}}}^{{\rm{an}}} \!\!=\!\!\left[\! {\frac{{2{n_{{\rm{an}}}}\! -\! {N_{{\rm{an}}}}}}{{{N_{{\rm{an}}}}}},\!\frac{{2{n_{{\rm{an}}}}\! - \!{N_{{\rm{an}}}} \!+ \!2}}{{{N_{{\rm{an}}}}}}} \!\right) ,
\end{equation}
\begin{equation}
{\mathcal{B}_{{\rm{de}}}} \!=\!\!\! \bigcup\limits_{{n_{{\rm{de}}}} = 0}^{{N_{{\rm{de}}}} - 1}\!\!\!\! {\Lambda _{{n_{{\rm{de}}}}}^{{\rm{de}}}} , \ \Lambda _{{n_{{\rm{de}}}}}^{{\rm{de}}}\! =\! \left[ {\frac{{{N_{\tau}}{n_{{\rm{de}}}}}}{{{N_{{\rm{de}}}}{N_{\rm{v}}}\Delta f}},\frac{{{N_{\tau}}\left( {{n_{{\rm{de}}}}\! +\! 1} \right)}}{{{N_{{\rm{de}}}}{N_{\rm{v}}}\Delta f}}} \right),
\end{equation}
\begin{equation}
{\mathcal{B}_{{\rm{do}}}}\!\!=\!\!\!\! \bigcup\limits_{{n_{{\rm{do}}}} = 0}^{{N_{{\rm{do}}}} - 1}\!\!\!\! {\Lambda _{{n_{{\rm{do}}}}}^{{\rm{do}}}} , \Lambda _{{n_{{\rm{do}}}}}^{{\rm{do}}} \!\!=\!\!\left[ \!{\frac{{{N_{\rm{d}}}\left( {{n_{{\rm{do}}}}\!\! -\! {\textstyle{{{N_{{\rm{do}}}}} \over 2}}} \right)}}{{{N_{{\rm{do}}}}N{T_{{\rm{sym}}}}}},\!\frac{{{N_{\rm{d}}}\left( {{n_{{\rm{do}}}}\!\! -\! {\textstyle{{{N_{{\rm{do}}}}} \over 2}}\! +\! 1} \right)}}{{{N_{{\rm{do}}}}N{T_{{\rm{sym}}}}}}}\! \right),
\end{equation}
\end{subequations}
where ${N_\tau } \buildrel \Delta \over =  {\frac{{{N_{\rm{v}}}{N_{\rm{g}}}}}{{{N_{\rm{c}}}}}} $, and $\frac{2}{{{N_{{\rm{an}}}}}}$, $ \frac{{{N_{\tau}}}}{{{N_{{\rm{de}}}}{N_{\rm{v}}}\Delta f}}$ and $ \frac{{{N_{\rm{d}}}}}{{{N_{{\rm{do}}}}N{T_{{\rm{sym}}}}}}$ are the intervals in directional cosine, delay and Doppler frequency, which can be flexibly adjusted by properly selecting ${N_{{\rm{an}}}} \! =\! \left\lceil {{F_{{\rm{an}}}}M\frac{{{f_{\rm{c}}}}}{{{f_{\rm{o}}}}}} \right\rceil $,\footnote{ For fixed ${f_{\rm{o}}}$, the angular resolution of the antenna array is proportional to ${f_{\rm{c}}}$.} ${N_{{\rm{de}}}} = {F_{{\rm{de}}}}{N_{\tau}}$, and ${N_{{\rm{do}}}} = {F_{{\rm{do}}}}{N_{\rm{d}}}$, where ${F_{{\rm{an}}}}$, ${F_{{\rm{de}}}}$, and ${F_{{\rm{do}}}}$  are referred to as fine factors.

Denote ${\Gamma _u} \buildrel \Delta \over = \big\{ \left( {{\Omega _{u,0}},{\tau _{u,0}},{\nu _{u,0}}} \right),\left( {{\Omega _{u,1}},{\tau _{u,1}},{\nu _{u,1}}} \right), \cdots ,$ $\left( {{\Omega _{u,{P_u} - 1}},{\tau _{u,{P_u} - 1}},{\nu _{u,{P_u} - 1}}} \right) \big\}$ as the path parameter set of UT $u$, and let  ${\Lambda _{{n_{{\rm{an}}}},{n_{{\rm{de}}}},{n_{{\rm{do}}}}}} \buildrel \Delta \over = \Lambda _{{n_{{\rm{an}}}}}^{{\rm{an}}} \times \Lambda _{{n_{{\rm{de}}}}}^{{\rm{de}}} \times \Lambda _{{n_{{\rm{do}}}}}^{{\rm{do}}}$. The SFT domain channel in (\ref{equ-chan2}) can be rewritten as
\begin{align}\label{equ-chan3}
{\bf{h}}_u^{{\rm{SFT}}} &= \sum\limits_{{n_{{\rm{an}}}} = 0}^{{N_{{\rm{an}}}} - 1} \sum\limits_{{n_{{\rm{de}}}} = 0}^{{N_{{\rm{de}}}} - 1} \sum\limits_{{n_{{\rm{do}}}} = {0}}^{{N_{{\rm{do}}}} - 1} \sum\limits_{\left( {{\Omega _{u,p}},{\tau _{u,p}},{\nu _{u,p}}} \right) \in {\Gamma _u} \cap {\Lambda _{{n_{{\rm{an}}}},{n_{{\rm{de}}}},{n_{{\rm{do}}}}}}}^{}\!\!\!\!\!\! \!\!\!{\gamma _{u,p}} \nonumber \\
&\qquad\qquad\qquad\qquad\qquad\ \times {\bf{p}}\left( {{\Omega _{u,p}},{\tau _{u,p}},{\nu _{u,p}}} \right)     .
\end{align}

We approximate the triple steering vector ${\bf{p}}\left( {{\Omega _{u,p}},\!{\tau _{u,p}},\!{\nu _{u,p}}} \right)$ for $\left( {{\Omega _{u,p}},\!{\tau _{u,p}},\!{\nu _{u,p}}} \right) \!\in \!{\Gamma _u} \cap {\Lambda _{{n_{{\rm{an}}}},{n_{{\rm{de}}}},{n_{{\rm{do}}}}}}$ as the sampled triple steering vector  ${\bf{p}}\left( {{\Omega _{{n_{{\rm{an}}}}}},{\tau _{{n_{{\rm{de}}}}}},{\nu _{{n_{{\rm{do}}}}}}} \right)$, where ${\Omega _{{n_{{\rm{an}}}}}} = \frac{{2{n_{{\rm{an}}}} - {N_{{\rm{an}}}}}}{{{N_{{\rm{an}}}}}}$, ${\tau _{{n_{{\rm{de}}}}}} = \frac{{{N_{\tau}}{n_{{\rm{de}}}}}}{{{N_{{\rm{de}}}}{N_{\rm{v}}}\Delta f}}$, and ${\nu _{{n_{{\rm{do}}}}}} \!=\! \frac{{{N_{\rm{d}}}\left( {{n_{{\rm{do}}}} - {{{N_{{\rm{do}}}}} \mathord{\left/
					{\vphantom {{{N_{{\rm{do}}}}} 2}} \right.
					\kern-\nulldelimiterspace} 2}} \right)}}{{{N_{{\rm{do}}}}N{T_{{\rm{sym}}}}}}$.
This approximation tends to be accurate as the numbers of divided subsets ${{N_{{\rm{an}}}}}$, ${{N_{{\rm{de}}}}}$ and ${{N_{{\rm{do}}}}}$ tend to be relatively large. 
We then define a matrix ${\bf{P}}$ that consists of sampled triple steering vectors and its $( {n_{{\rm{do}}}}{N_{{\rm{an}}}}{N_{{\rm{de}}}}+ {n_{{\rm{de}}}}{N_{{\rm{an}}}} + {n_{{\rm{an}}}} )$-th column is ${\bf{p}}\left( {{\Omega _{{n_{{\rm{an}}}}}},{\tau _{{n_{{\rm{de}}}}}},{\nu _{{n_{{\rm{do}}}}}}} \right)$, which represents the triple steering vector pointing towards the sampled
directional cosine ${\Omega _{{n_{{\rm{an}}}}}}$, delay ${\tau _{{n_{{\rm{de}}}}}}$ and Doppler frequency ${\nu _{{n_{{\rm{do}}}}}}$. 
Moreover, according to (\ref{equ-tb-vector}), $\bf{P}$ can be expressed as
\begin{equation}\label{equ-P-stru}
{\bf{P}} \buildrel \Delta \over = {\bf{D}} \otimes \left( {{\bf{V}}\left( {{\bf{U}} \otimes {{\bf{I}}_{{N_{{\rm{an}}}}}}} \right)} \right) \in {\mathbb{C}^{M{N_{\rm{v}}}N \times {N_{{\rm{an}}}}{N_{{\rm{de}}}}{N_{{\rm{do}}}}}},
\end{equation}
where ${\bf{D}} \in {\mathbb{C}^{N \times {N_{{\rm{do}}}}}}$ with ${\left[ {\bf{D}} \right]_{i,j}} = {e^{{\bar \jmath}2\pi i\frac{{{N_{\rm{d}}}\left( {j - {{{N_{{\rm{do}}}}} \mathord{\left/
							{\vphantom {{{N_{{\rm{do}}}}} 2}} \right.
							\kern-\nulldelimiterspace} 2}} \right)}}{{{N_{{\rm{do}}}}N}}}}$, 
${\bf{U}} \in {\mathbb{C}^{{N_{\rm{v}}} \times {N_{{\rm{de}}}}}}$ with ${\left[ {\bf{U}} \right]_{i,j}} = {e^{ - {\bar \jmath}2\pi \left( {i + {k_0}} \right)\frac{{{N_{\tau}}j}}{{{N_{{\rm{de}}}}{N_{\rm{v}}}}}}}$, 
${\bf{V}} = {\rm{diag}}\left\{ {{{\bf{V}}_{{k_0}}}, \cdots ,{{\bf{V}}_{{k_{{N_{\rm{v}}} - 1}}}}} \right\}$, and ${{\bf{V}}_k} \in {\mathbb{C}^{M \times {N_{{\rm{an}}}}}}$ with ${\left[ {{{\bf{V}}_k}} \right]_{i,j}} = {e^{ - {\bar \jmath}2\pi \left( {{f_{\rm{c}}} + k\Delta f} \right)i\Delta \tau \frac{{2j - {N_{{\rm{an}}}}}}{{{N_{{\rm{an}}}}}}}}$.
Specially, when the spatial-wideband effect is not considered, ${{\bf{V}}_k}$ will be reduced to ${\bf{\tilde V}} \in {\mathbb{C}^{M \times {N_{{\rm{an}}}}}}$ with ${[ {{\bf{\tilde V}}} ]_{i,j}} = {e^{ - {\bar \jmath}2\pi {f_{\rm{c}}}i\Delta \tau \frac{{2j - {N_{{\rm{an}}}}}}{{{N_{{\rm{an}}}}}}}}$, which is independent of the subcarrier, and (\ref{equ-P-stru}) can be rewritten as ${\bf{P}} = {\bf{D}} \otimes {\bf{U}} \otimes {\bf{\tilde V}}$ in this case. 

Then the SFT domain channel in (\ref{equ-chan3}) can be approximated as
\begin{equation}\label{equ-chan-sft}
{\bf{h}}_u^{{\rm{SFT}}} = {\bf{Ph}}_u^{{\rm{TB}}},
\end{equation}
where ${{\bf{h}}_u^{{\rm{TB}}}}\in {\mathbb{C}^{{N_{{\rm{an}}}}{N_{{\rm{de}}}}{N_{{\rm{do}}}} \times 1}}$ can be expressed as
\begin{equation}\label{equ_h_bdd}
{\left[ {{\bf{h}}_u^{{\rm{TB}}}} \right]_{ {{n_{{\rm{do}}}}{N_{{\rm{an}}}}{N_{{\rm{de}}}} + {n_{{\rm{de}}}}{N_{{\rm{an}}}} + {n_{{\rm{an}}}}}}} \!\!\buildrel \Delta \over =\!\!\!\! \sum\limits_{\left( {{\Omega _{u,p}},{\tau _{u,p}},{\nu _{u,p}}} \right) \in {\Gamma _u} \cap {\Lambda _{{n_{{\rm{an}}}},{n_{{\rm{de}}}},{n_{{\rm{do}}}}}}}^{}\!\!\!\!\!\!\!\!\!\!\!\! {{\gamma _{u,p}}} .
\end{equation}
In the channel representation of (\ref{equ-chan-sft}), all the UTs share the same set of sampled triple steering vectors. Each sampled triple steering vector corresponds to a physical triple-beam in the SFT domain. Therefore, we refer to it as the \emph{triple-beam (TB) based channel model}. ${{\bf{h}}_u^{{\rm{TB}}}}$ is the TB domain channel vector and $\bf{P}$ is referred to as the TB matrix. Furthermore, 
the SFT domain channel covariance matrix of UT $u$ can be given by
\begin{align}\label{equ-Rsft}
{\bf{R}}_u^{{\rm{SFT}}}& = \mathbb{E}\left\{ {{\bf{h}}_u^{{\rm{SFT}}}{{\left( {{\bf{h}}_u^{{\rm{SFT}}}} \right)}^{\rm{H}}}} \right\} = {\bf{P}}\mathbb{E}\left\{ {{\bf{h}}_u^{{\rm{TB}}}{{\left( {{\bf{h}}_u^{{\rm{TB}}}} \right)}^{\rm{H}}}} \right\}{\bf{P}}^{\rm{H}} \nonumber \\
&= {\bf{PR}}_u^{{\rm{TB}}}{\bf{P}}^{\rm{H}} \in {\mathbb{C}^{M{N_{\rm{v}}}N \times M{N_{\rm{v}}}N}},
\end{align}
where ${\bf{R}}_u^{{\rm{TB}}} \!\!\buildrel \Delta \over = \mathbb{E}\left\{ {{\bf{h}}_u^{{\rm{TB}}}{{\left( {{\bf{h}}_u^{{\rm{TB}}}} \right)}^{\rm{H}}}} \right\}\!\in \!{\mathbb{C}^{{N_{{\rm{an}}}}{N_{{\rm{de}}}}{N_{{\rm{do}}}} \times {N_{{\rm{an}}}}{N_{{\rm{de}}}}{N_{{\rm{do}}}}}}$ is the TB domain channel covariance matrix of UT $u$, also referred to as the TB domain statistical CSI. 
We assume that TB domain channel coefficients follow independent complex Gaussian distributions with zero mean and different variances. Hence, ${\bf{R}}_u^{{\rm{TB}}}$ is a diagonal matrix with the $\left( {{n_{{\rm{do}}}}{N_{{\rm{an}}}}{N_{{\rm{de}}}}\! +\! {n_{{\rm{de}}}}{N_{{\rm{an}}}}\! +\! {n_{{\rm{an}}}}\!} \right)$-th diagonal element expressed as $\sum\nolimits_{\left( {{\Omega _{u,p}},{\tau _{u,p}},{\nu _{u,p}}} \right) \in {\Gamma _u} \cap {\Lambda _{{n_{{\rm{an}}}},{n_{{\rm{de}}}},{n_{{\rm{do}}}}}}} {\beta _{u,p}^2} $.
Due to the limited propagation path number, Doppler spread, and small angle spread, the TB domain channel typically exhibits sparsity in HF skywave massive MIMO-OFDM communication. 
Hence, compared with ${\bf{R}}_u^{{\rm{SFT}}}$, the number of non-zero elements in ${\bf{R}}_u^{{\rm{TB}}}$ is substantially smaller since it is a diagonal matrix and most of the elements on the diagonal  are approximately zero. Moreover, ${\bf{R}}_u^{{\rm{TB}}}$ varies slowly relative to the communication timescale \cite{8354789,9440710,7332961}. Therefore, there are sufficient resources to acquire ${\bf{R}}_u^{{\rm{TB}}}$. For instance, the low-complexity method in \cite{LAA} can estimate the spatial-frequency beam domain statistical CSI directly without involving the instantaneous channel, which can be straightforwardly extended to the TB based channels. Thus we assume that the TB domain statistical CSI ${\bf{R}}_u^{{\rm{TB}}}$ is available at the BS in the rest of the paper.

In our proposed TB based channel model, we generalize the concept of the spatial-beam in massive MIMO \cite{7045498,yxl} to the TB consisting of the spatial-beam, the frequency-beam and the temporal-beam. Moreover, when ${f_{\rm{o}}} = {f_{\rm{c}}}$, ${N_{{\rm{an}}}} = M$, ${N_{{\rm{de}}}} = {N_{\tau}}$, ${N_{{\rm{do}}}} = {N_{\rm{d}}}$, the spatial-wideband effect is ignored, the columns of $\bf{P}$ will become orthogonal. Besides, we can set ${N_{{\rm{an}}}}> M$, ${N_{{\rm{de}}}} > {N_{\tau}}$, and ${N_{{\rm{do}}}} > {N_{\rm{d}}}$ to make sampling intervals in (\ref{equ-sam_int}) smaller and obtain a higher resolution of the angle, delay, and Doppler frequency, respectively. Although the orthogonality among the columns in $\bf{P}$ no longer holds under such a setting, it makes the channel model more accurate and ensures more accurate channel acquisition. 
Moreover, the TB based channel model can also be applied in generic massive MIMO communications and  wireless channels with different carrier frequencies, different propagation scenarios, and different system configurations although we focus on the HF skywave channels in this work.

\section{Channel acquisition and pilot design}
In this section, we first investigate optimal channel estimation for pilot segments. Then we find the conditions to minimize the NMSE of the channel estimate and develop a pilot design,  including UT grouping and pilot scheduling. Moreover, we propose a method to predict the channel at  data segments based on the estimated TB domain channel.

\subsection{Channel Estimation for Pilot Segment}

We first consider the estimation of the SFT domain channel at pilot segments. According to (\ref{equ-chan-sft}) and the frame structure shown in Fig. 1, the SFT domain channel at pilot segments can be expressed as
\begin{equation}\label{equ-chan-sftp}
{\bf{h}}_u^{{\rm{SFT,p}}} = {\bf{\tilde P}}{\bf h}_u^{{\rm{TB}}} \in {\mathbb{C}^{M{N_{\rm{v}}}{N_{\rm{F}}} \times 1}},
\end{equation}
where
\begin{align}\label{equ-TB-til}
{\bf{\tilde P}} &\buildrel \Delta \over = \left( {{\bf{\Theta }} \otimes {{\bf{I}}_{{N_{\rm{v}}}}} \otimes {{\bf{I}}_M}} \right){\bf{P}} \nonumber \\
&= {\bf{\tilde D}} \otimes \left( {{\bf{V}}\left( {{\bf{U}} \otimes {{\bf{I}}_{{N_{{\rm{an}}}}}}} \right)} \right)\in {\mathbb{C}^{M{N_{\rm{v}}}{N_{\rm{F}}} \times {N_{{\rm{an}}}}{N_{{\rm{de}}}}{N_{{\rm{do}}}}}},
\end{align} 
${\bf{\Theta }} \in {\mathbb{C}^{{N_{\rm{F}}} \times N}}$ with the ${n_{\rm{F}}}$-th row being the $\left({{n_{\rm{F}}}{N_{\rm{S}}}+{n_{\rm{p}}}}\right)$-th row of ${{\bf{I}}_N}$, and ${\bf{\tilde D}} = {\bf{\Theta D}} \in {\mathbb{C}^{{N_{\rm{F}}} \times {N_{{\rm{do}}}}}}$.

In each timeslot in Fig. 1, the channel estimation should be performed to obtain the CSI using the pilot. 
We combine the current timeslot with the previous ${N_{\rm{F}}}-1$ timeslots to form a complete frame as in Fig. 1, where the current timeslot refers to the last timeslot in the whole frame.  In this case, we need to save the received signals at pilot segments in the previous ${N_{\rm{F}}}-1$ timeslots and perform channel estimation together with the received signal at the pilot segment in the current timeslot.

Let ${{\bf{x}}_u^{\rm{p}}} \in {\mathbb{C}^{{N_{\rm{v}}} \times 1}}$ denote the transmitted pilot of UT $u$ at valid subcarriers. The received signal ${\bf{y}} \in {\mathbb{C}^{M{N_{\rm{v}}}{N_{\rm{F}}} \times 1}}$ at the BS can be given by
\begin{equation}\label{equ-UL}
{\bf{y}} = \sum\limits_{u = 0}^{U - 1} {{{\bf{X}}_u}{\bf{h}}_u^{{\rm{SFT,p}}} + {\bf{z}}}  = {\bf{X}}{{\bf{h}}^{{\rm{SFT,p}}}} + {\bf{z}},
\end{equation}  
where ${{\bf{h}}^{{\rm{SFT,p}}}} \buildrel \Delta \over = {\left[ {{{\left( {{\bf{h}}_0^{{\rm{SFT,p}}}} \right)}^{\!\rm{T}}}\!,\!{{\left( {{\bf{h}}_1^{{\rm{SFT,p}}}} \right)}^{\!\rm{T}}}, \cdots ,{{\left( {{\bf{h}}_{U - 1}^{{\rm{SFT,p}}}} \right)}^{\!\rm{T}}}} \right]^{\!\rm{T}}} \!\!\!\!\!\in\! {\mathbb{C}^{M{N_{\rm{v}}}{N_{\rm{F}}}U \times 1}}$, ${{\bf{X}}_u} \!\!\buildrel \Delta \over =\! {{\bf{I}}_{N_{\rm{F}}}}\! \otimes ( {\rm{diag}}\!\left\{ {{\bf{x}}_u^{\rm{p}}} \right\} \otimes {{\bf{I}}_M} \!) \!\!\in\! {\mathbb{C}^{M{N_{\rm{v}}}{N_{\rm{F}}}\! \times\! M{N_{\rm{v}}}{N_{\rm{F}}}}}$, ${\bf{X}} \buildrel \Delta \over = \left[ {{{\bf{X}}_0},{{\bf{X}}_1}, \cdots ,{{\bf{X}}_{U - 1}}} \right] \in {\mathbb{C}^{M{N_{\rm{v}}}{N_{\rm{F}}} \times M{N_{\rm{v}}}{N_{\rm{F}}}U}}$, and ${\bf{z}} \in {\mathbb{C}^{M{N_{\rm{v}}}{N_{\rm{F}}} \times 1}}$ is the additive white Gaussian noise with distribution $\mathcal{CN}\left( {{\bf{z}};{\bf{0}},\sigma _{\rm{z}}^2{{\bf{I}}_{M{N_{\rm{v}}}{N_{\rm{F}}}}}} \right)$.  
Substituting (\ref{equ-chan-sftp}) into (\ref{equ-UL}), we can obtain that 
\begin{equation}\label{equ-cs}
{\bf{y}} = {\bf{X}}\left( {{{\bf{I}}_U} \otimes {\bf{\tilde P}}} \right){{\bf{h}}^{{\rm{TB}}}} + {\bf{z}} = {\bf{A}}{{\bf{h}}^{{\rm{TB}}}} + {\bf{z}},
\end{equation}
where ${{\bf{h}}^{{\rm{TB}}}} \!\buildrel \Delta \over = \!\!{\left[\! {{{\left( {{\bf{h}}_0^{{\rm{TB}}}} \right)}^{\!\rm{T}}}\!,\!{{\left( {{\bf{h}}_1^{{\rm{TB}}}} \right)}^{\!\rm{T}}}\!, \!\cdots \!,\!{{\left( {{\bf{h}}_{U - 1}^{{\rm{TB}}}} \right)}^{\!\rm{T}}}} \right]^{\!\rm{T}}}\!\!\!\!\! \in\!{\mathbb{C}^{{N_{{\rm{an}}}}{N_{{\rm{de}}}}{N_{{\rm{do}}}}U\! \times\! 1}}$ and ${\bf{A}} \buildrel \Delta \over = {\bf{X}}\left( {{{\bf{I}}_U} \otimes {\bf{\tilde P}}} \right)$. 

According to the relationship between the SFT domain channel and the TB domain channel shown in (\ref{equ-chan-sftp}), the estimation of ${\bf{h}}^{{\rm{SFT,p}}}$ can be converted to the estimation of ${\bf{h}}^{{\rm{TB}}}$.
Since the TB domain channel ${\bf{h}}^{{\rm{TB}}}$ is distributed as the complex Gaussian distribution, the optimal estimate of ${{\bf{h}}^{{\rm{TB}}}}$ is the MMSE estimate, which is given by
\begin{equation}\label{equ-MMSE}
{{\bf{\hat h}}^{{\rm{TB}}}} = {{\bf{R}}^{{\rm{TB}}}}{{\bf{A}}^{\rm{H}}}{\left( {{\bf{A}}{{\bf{R}}^{{\rm{TB}}}}{{\bf{A}}^{\rm{H}}} + \sigma _{\rm{z}}^2{{\bf{I}}_{M{N_{\rm{v}}}{N_{\rm{F}}}}}} \right)^{ - 1}}{\bf{y}},
\end{equation}
where ${{\bf{R}}^{{\rm{TB}}}}\! =\! \mathbb{E}\left\{\! {{{\bf{h}}^{{\rm{TB}}}}{{\left( {{{\bf{h}}^{{\rm{TB}}}}} \right)}^{\!\rm{H}}}} \!\right\} \!= \!{\rm{diag}}\left\{\! {{\bf{R}}_0^{{\rm{TB}}}\!,\!{\bf{R}}_1^{{\rm{TB}}}\!, \!\cdots \!,\!{\bf{R}}_{U - 1}^{{\rm{TB}}}} \right\}$. The estimate of ${{\bf{ h}}^{{\rm{SFT,p}}}}$ can be expressed as
\begin{align}\label{equ_sft_mmse}
{{\bf{\hat h}}^{{\rm{SFT,p}}}} &= \left( {{{\bf{I}}_U} \otimes {\bf{\tilde P}}} \right){{\bf{\hat h}}^{{\rm{TB}}}} \nonumber \\
&= {{\bf{R}}^{{\rm{SFT,p}}}}{{\bf{X}}^{\rm{H}}}{\left( {{\bf{X}}{{\bf{R}}^{{\rm{SFT,p}}}}{{\bf{X}}^{\rm{H}}} \!+\! \sigma _{\rm{z}}^2{{\bf{I}}_{M{N_{\rm{v}}}{N_{\rm{F}}}}}}\! \right)^{ - 1}}{\bf{y}},
\end{align}
where 
\begin{align}
{{\bf{R}}^{{\rm{SFT,p}}}} &= \left( {{{\bf{I}}_U} \otimes {\bf{\tilde P}}} \right){{\bf{R}}^{{\rm{TB}}}}{\left( {{{\bf{I}}_U} \otimes {\bf{\tilde P}}} \right)^{\rm{H}}} \nonumber\\
&= \mathbb{E}\left\{ {{{\bf{h}}^{{\rm{SFT,p}}}}{{\left( {{{\bf{h}}^{{\rm{SFT,p}}}}} \right)}^{\rm{H}}}} \right\} \nonumber \\
&= {\rm{diag}}\left\{ {{\bf{R}}_0^{{\rm{SFT,p}}},{\bf{R}}_1^{{\rm{SFT,p}}}, \cdots ,{\bf{R}}_{U - 1}^{{\rm{SFT,p}}}} \right\}.
\end{align}
Note that (\ref{equ_sft_mmse}) is also the MMSE estimate of ${{\bf{ h}}^{{\rm{SFT,p}}}}$.

As an important measure for channel estimation, the NMSE of ${{\bf{\hat h}}^{{\rm{SFT,p}}}}$ is defined as
\begin{align}\label{equ-nmse-ini}
&{\rm{NMSE}} \buildrel \Delta \over =\! \frac{{\mathbb{E}\left\{ {{{\left\| {{{\bf{h}}^{{\rm{SFT,p}}}} - {{{\bf{\hat h}}}^{{\rm{SFT,p}}}}} \right\|}^2}} \right\}}}{{\mathbb{E}\left\{ {{{\left\| {{{\bf{h}}^{{\rm{SFT,p}}}}} \right\|}^2}} \right\}}} \nonumber \\
&= \!\!\sum\limits_{u = 0}^{U - 1}\!\! {\frac{1}{{M{N_{\rm{v}}}{N_{\rm{F}}}U{\vartheta _u}}}{\rm{tr}}\!\left\{\! {{\bf{R}}_u^{{\rm{SFT,p}}} \!\!-\! {\bf{R}}_u^{{\rm{SFT,p}}}{\bf{X}}_u^{\rm{H}}{{\bf{C}}^{ - 1}}{{\bf{X}}_u}{\bf{R}}_u^{{\rm{SFT,p}}}} \!\right\}} ,
\end{align}
where ${\bf{C}} \buildrel \Delta \over = \sum\nolimits_{u = 0}^{U - 1} {{{\bf{X}}_u}{\bf{R}}_u^{{\rm{SFT,p}}}{\bf{X}}_u^{\rm{H}}}  + \sigma _{\rm{z}}^2{{\bf{I}}_{M{N_{\rm{v}}}{N_{\rm{F}}}}}$ and ${\vartheta _u} \buildrel \Delta \over = \sum\nolimits_{p = 0}^{\!{P_u} \!-\! 1} \!\!{\beta _{u,p}^2}\! $ is the large-scale fading between the BS and UT $u$.

From (\ref{equ-nmse-ini}), the NMSE of ${{\bf{\hat h}}^{{\rm{SFT,p}}}}$ can be regarded as the average of the NMSE of each UT. For UT $u$, the term $\sum\limits_{u = 0}^{U - 1} \!{{{\bf{X}}_u}{\bf{R}}_u^{{\rm{SFT,p}}}{\bf{X}}_u^{\rm{H}}} $ in ${\bf{C}}$ implies that the NMSE of each UT is related to not only its own statistical CSI but also that of other UTs, which is referred to as the inter-UT interference.

\subsection{Pilot Design}

We next discuss the pilot design for the optimal channel estimation. 
Motivated by the widely used phase shift pilots \cite{8654199}, the transmitted pilot of UT $u$ is given by
\begin{equation}\label{equ-pilot}
{\bf{x}}_u^{\rm{P}}\!\! =\!\! {\sigma _{\rm{p}}}{\bf{x}}_{\rm{c}} \circ{\big[ {{e^{\! - {\bar \jmath}2\pi {k_0}\!\frac{{{N_{\tau}}{\phi _u}}}{{{N_{{\rm{de}}}}{N_{\rm{v}}}}}}}\!,\!{e^{ \!- {\bar \jmath}2\pi {k_1}\!\frac{{{N_{\tau}}{\phi _u}}}{{{N_{{\rm{de}}}}{N_{\rm{v}}}}}}}\!, \!\cdots \!,\!{e^{\! - {\bar \jmath}2\pi {k_{{N_{\rm{v}}}\! -\! 1}}\!\frac{{{N_{\tau}}{\phi _u}}}{{{N_{{\rm{de}}}}{N_{\rm{v}}}}}}}} \big]^{\!\rm{T}}}\!,
\end{equation}
where $\sigma _{\rm{p}}$ is the square root of the pilot transmit power,
${\phi _u} \!\in\! \left\{ {0,{N_{{\rm{de}}}}, \cdots ,\left( {\left\lfloor {{{{N_{\rm{v}}}} \mathord{\left/
					{\vphantom {{{N_{\rm{v}}}} {{N_{\tau}}}}} \right.
					\kern-\nulldelimiterspace} {{N_{\tau}}}}} \right\rfloor  - 1} \right){N_{{\rm{de}}}}} \right\}$ 
is referred to as the phase shift factor
and ${\bf{x}}_{\rm{c}}$ is a sequence with unity modulus elements. A good choice for ${\bf{x}}_{\rm{c}}$ is the Zadoff-Chu (ZC) sequence \cite{1054840}, which has been extensively used in the fifth generation (5G) wireless networks \cite{dahlman20205g}.  
Note that ${\bf{\tilde P}}$ in (\ref{equ-TB-til}) can be rewritten as
\begin{equation}
{\bf{\tilde P}}  = {\bf{\bar P}}\left( {{{\bf{I}}_{{N_{{\rm{do}}}}}} \otimes {{\bf{I}}_{\left\lfloor {{{{N_{\rm{v}}}} \mathord{\left/
						{\vphantom {{{N_{\rm{v}}}} {{N_{\tau}}}}} \right.
						\kern-\nulldelimiterspace} {{N_{\tau}}}}} \right\rfloor {N_{{\rm{de}}}} \times {N_{{\rm{de}}}}}} \otimes {{\bf{I}}_{{N_{{\rm{an}}}}}}} \right),
\end{equation}	
where
\begin{equation}\label{equ-stru-P}
{\bf{\bar P}} \!\buildrel \Delta \over =\! {\bf{\tilde D}} \otimes\! \left( {{\bf{V}}\!\left( {{\bf{\bar U}} \!\otimes\! {{\bf{I}}_{{N_{{\rm{an}}}}}}} \right)} \right) \!\in\! {\mathbb{C}^{M{N_{\rm{v}}}{N_{\rm{F}}} \!\times\! {N_{{\rm{an}}}}\left\lfloor {{{{N_{\rm{v}}}} \mathord{\left/
					{\vphantom {{{N_{\rm{v}}}} {{N_{\tau}}}}} \right.
					\kern-\nulldelimiterspace} {{N_{\tau}}}}} \right\rfloor {N_{{\rm{de}}}}{N_{{\rm{do}}}}}},
\end{equation}
and ${\bf{\bar U}} \in {\mathbb{C}^{{N_{\rm{v}}} \times \left\lfloor {{{{N_{\rm{v}}}} \mathord{\left/
					{\vphantom {{{N_{\rm{v}}}} {{N_{\tau}}}}} \right.
					\kern-\nulldelimiterspace} {{N_{\tau}}}}} \right\rfloor {N_{{\rm{de}}}}}}$ with ${\left[ {{\bf{\bar U}}} \right]_{i,j}}\! =\! {e^{ - {\bar \jmath}2\pi \left( {i + {k_0}} \right)\frac{{{N_{\tau}}j}}{{{N_{{\rm{de}}}}{N_{\rm{v}}}}}}}$.
It can be checked that ${{\bf{X}}_u}{\bf{\tilde P}} = {\sigma _{\rm{p}}}{{\bf{X}}_{\rm{c}}}{\bf{\bar P}}{{\bf{S}}_u}$, where ${{\bf{X}}_{\rm{c}}} \buildrel \Delta \over = {{\bf{I}}_{{N_{\rm{F}}}}} \otimes \left( {{\rm{diag}}\left\{ {{{\bf{x}}_{\rm{c}}}} \right\} \otimes {{\bf{I}}_M}} \right)$ and 
\begin{equation}
{{\bf{S}}_u} \!\buildrel \Delta \over= \!\big( {{{\bf{I}}_{{N_{{\rm{do}}}}}} \!\!\otimes \!{{\left[ {{{\bf{0}}_{{N_{{\rm{de}}}} \!\times\! {\phi _u}}},\!{{\bf{I}}_{{N_{{\rm{de}}}}}},\!{{\bf{0}}_{{N_{{\rm{de}}}} \!\times\! \left(\! {\left( {\left\lfloor {{{{N_{\rm{v}}}} \mathord{\left/
											{\vphantom {{{N_{\rm{v}}}} {{N_{\tau}}}}} \right.
											\kern-\nulldelimiterspace} {{N_{\tau}}}}} \right\rfloor  \!- \!1} \right){N_{{\rm{de}}}} \!-\! {\phi _u}} \!\right)}}} \right]}^{\!\rm{T}}} \!\!\!\otimes\! {{\bf{I}}_{{N_{{\rm{an}}}}}}}\! \big) 
\end{equation}
is the selection matrix depending on ${\phi _u}$. Thus $\bf{A}$ can be rewritten as
\begin{equation}\label{equ-stru-A}
{\bf{A}} = {\sigma _{\rm{p}}}{{\bf{X}}_{\rm{c}}}{\bf{\bar P}}\left[ {{{\bf{S}}_0}, \cdots ,{{\bf{S}}_{U - 1}}} \right].
\end{equation}

Next, we properly allocate all pilot sequences to each UT and optimize the channel estimation performance. To this end, we have the following theorem, proved in Appendix A, which provides the criterion for the UT grouping and  pilot scheduling.

\noindent {\bf Theorem 1}:  When $M,{N_{\rm{v}}},{N_{\rm{F}}} \to \infty $, the minimum value of NMSE is given by
\begin{align}\label{equ-nmse_min}
&{\rm{NMS}}{{\rm{E}}_{\min }}\! = \sum\limits_{u = 0}^{U - 1} \frac{1}{{M{N_{\rm{v}}}{N_{\rm{F}}}U{\vartheta _u}}}{\rm{tr}}\Big\{ {\bf{R}}_u^{{\rm{SFT,p}}}   \nonumber\\ 
&\quad\quad\quad\  -\!{\bf{R}}_u^{{\rm{SFT,p}}}{{\left( {{\bf{R}}_u^{{\rm{SFT,p}}}\! +\! \frac{{\sigma _{\rm{z}}^2}}{{\sigma _{\rm{p}}^2}}{{\bf{I}}_{M{N_{\rm{v}}}{N_{\rm{F}}}}}} \right)}^{ - 1}}\!\!{\bf{R}}_u^{{\rm{SFT,p}}} \!\Big\} , 
\end{align}
provided that, for $\forall u,u'$ and $u \ne u'$, one of the following conditions is satisfied
\begin{enumerate}[1)]
	\item ${\phi _u} \ne {\phi _{u'}},\ {\rm{ }}{\bf{R}}_u^{{\rm{TB}}}{\bf{R}}_{u'}^{{\rm{TB}}} \ne {\bf{0}}$,
	\item ${\bf{R}}_u^{{\rm{TB}}}{\bf{R}}_{u'}^{{\rm{TB}}} = {\bf{0}}$.
\end{enumerate}

From Theorem 1, when the NMSE achieves the minimum value, the NMSE of the channel estimate of each UT is only related to its own statistical CSI and is not related to other UTs, which means that the inter-UT interference is eliminated. 
Furthermore, ${\bf{R}}_u^{{\rm{TB}}}{\bf{R}}_{u'}^{{\rm{TB}}} \ne {\bf{0}}$ means that there is an overlap between the TB domain channels of UTs $u$ and $u'$ while ${\bf{R}}_u^{{\rm{TB}}}{\bf{R}}_{u'}^{{\rm{TB}}} = {\bf{0}}$ implies that the TB domain channels of UTs $u$ and $u'$ are non-overlapping.
Therefore, condition 1) implies that to minimize the NMSE, UTs with overlapping TB domain channels should be allocated pilot sequences with different phase shift factors while condition 2) implies that UTs with non-overlapping TB domain channels can reuse the same pilot sequence.
Such a result is intuitive because we need pilot sequences with different pilot shift factors to separate UTs with overlapping TB domain channels along the delay-dimension at the BS while UTs with non-overlapping TB domain channels can be directly separated at the BS, thus the same pilot sequence can be reused to reduce pilot overhead  and increases the number of UTs that can be served.
The above result is also consistent with findings in conventional massive MIMO literature \cite{7045498,7524027}.

Next, we define the degree of channel overlap between UTs $u$ and $u'$ as
\begin{equation}\label{channeloverlap-bdd}
{\rho _{u,u'}} \buildrel \Delta \over = \frac{{{\rm{tr}}\left\{ {{\bf{R}}_u^{{\rm{TB}}}{\bf{R}}_{u'}^{{\rm{TB}}}} \right\}}}{{{{\left\| {{\bf{R}}_u^{{\rm{TB}}}} \right\|}_{\rm{F}}}{{\left\| {{\bf{R}}_{u'}^{{\rm{TB}}}} \right\|}_{\rm{F}}}}}.
\end{equation}
It can be directly checked that $0 \le {\rho _{u,u'}} \le 1$.
In practical systems, the conditions in Theorem 1 are hard to be well satisfied due to limited $M$, ${N_{\rm{v}}}$, and ${N_{\rm{F}}}$. However,  the power of the TB domain channel of each UT is typically concentrated on a narrow support and the channel overlap degree among UTs  with different supports of TB domain channel can be small.
Therefore, we can divide all UTs into $S$ groups 
($S \le {\left\lfloor {{{{N_{\rm{v}}}} \mathord{\left/{\vphantom {{{N_{\rm{v}}}} {{N_{\tau}}}}} \right.\kern-\nulldelimiterspace} {{N_{\tau}}}}} \right\rfloor } $) according to the following UT grouping criterion.
\begin{enumerate}[i).]
	\item The degree of channel overlap between any two UTs in the same group should be as low as possible.
	\item UTs with a high degree of channel overlap should be allocated to different groups.
\end{enumerate}
Meanwhile, we allocate pilot sequences with different phase shift factors to each UT group. With UT grouping and pilot scheduling, the inter-UT interference can be suppressed and the channel estimation performance can be improved.

Motivated by the hierarchical clustering \cite{8853526}, we propose a UT grouping algorithm shown in Algorithm 1 according to the UT grouping criterion. Specifically, we first place each UT in a group of its own. Then in each iteration, we search for a pair of UT groups, where the average channel overlap degree between UTs in these two groups is the smallest, and combine them into one UT group. The iteration is terminated when the current number of groups is equal to $S$.

\IncMargin{0.2em}
{\LinesNumberedHidden
\begin{algorithm}
	\SetAlgoNoLine 
	\caption{UT grouping algorithm}
	\KwIn{
		The number of UTs $U$; the number of groups to divide $S$; the degree of overlap between UTs $\left\{ {{\rho _{u,u'}},u,u' = 0,1, \cdots ,U - 1} \right\}$\\
	}
	
	{\bfseries Initialize:} The index set of UT groups $\Psi  = \{ 0,1, \cdots ,U - 1 \}$; the initial UT grouping  result ${\Upsilon _s} = \left\{ s \right\},s \in \Psi $\\
	
	\While{ $\left| \Psi  \right| > S  $}{
		$\left\{ {{s_1},{s_2}} \right\} = \mathop {\arg \min }\limits_{s \in \Psi ,s' \in \Psi \backslash s} \sum\limits_{u \in {\Upsilon _s},u' \in {\Upsilon _{s'}}}^{} {\frac{{{\rho _{u,u'}}}}{{\left| {{\Upsilon _s}} \right|\left| {{\Upsilon _{s'}}} \right|}}} $\\
		${\Upsilon _{{s_1}}} \leftarrow {\Upsilon _{{s_1}}} \cup {\Upsilon _{{s_2}}}$ \\
		$\Psi  \leftarrow \Psi \backslash {s_2}$ \\
				
	}

	\KwOut{
		UT grouping result $\left\{ {{\Upsilon _s},s \in \Psi} \right\}$\\
	}
	
\end{algorithm}
}

Note that the TB domain statistical CSI includes three dimensions, i.e., the spatial-beam dimension, the frequency-beam dimension, and the temporal-beam dimension. In the above discussion, all these three dimensions are utilized for UT grouping and the computational complexity of ${\rho _{u,u'}}$ in (\ref{channeloverlap-bdd}) is $\mathcal{O}\left( {{N_{{\rm{an}}}}{N_{{\rm{de}}}}{N_{{\rm{do}}}}} \right)$. Typically, the angle spread of the HF skywave channel is small \cite{8904116,balser1962some,HFcom}, which implies that the TB domain channel of many UTs can be non-overlapping along the spatial-beam dimension, and we can just use the spatial-beam dimension of the TB domain statistical CSI to perform UT grouping. To this end, we first define 
\begin{equation}
	{\bf{R}}_u^{\rm{B}}\!\buildrel \Delta \over = \!\!\!\sum\limits_{i = 0}^{{N_{{\rm{de}}}}{N_{{\rm{do}}}} - 1} \!{\sum\limits_{j = 0}^{{N_{{\rm{de}}}}{N_{{\rm{do}}}} - 1} \!\!\!\!\!\!\!\!{{{\left[ {{\bf{R}}_u^{{\rm{TB}}}} \right]}_{i{N_{{\rm{an}}}}:\left( {i + 1} \right){N_{{\rm{an}}}} - 1,\hfill\atop
					{jN_{{\rm{an}}}}:\left( {j + 1} \right){N_{{\rm{an}}}} - 1\hfill}}} }  \!\!\in\! {\mathbb{C}^{{N_{{\rm{an}}}} \!\times\! {N_{{\rm{an}}}}}}.
\end{equation}
Then we have the following theorem, proved in Appendix B.

\noindent {\bf Theorem 2}: For arbitrary ${N_{\rm{v}}}$ and ${N_{\rm{F}}}$,  when $M \to \infty $, the  NMSE in (\ref{equ-nmse-ini}) can be rewritten as
\begin{align}\label{equ-theorem2}
	&{\rm{NMSE}} = \sum\limits_{u = 0}^{U - 1} \frac{1}{{M{N_{\rm{v}}}{N_{\rm{F}}}U{\vartheta _u}}}{\rm{tr}}\big\{ {\bf{R}}_u^{{\rm{SFT,p}}} \nonumber \\
	&\quad\quad\quad\quad\quad\quad\quad\quad\ - {\bf{R}}_u^{{\rm{SFT,p}}}{\bf{X}}_u^{\rm{H}}{{{\bf{\bar C}}}_u^{ - 1}}{{\bf{X}}_u}{\bf{R}}_u^{{\rm{SFT,p}}} \big\} ,
\end{align}
where
\begin{equation}
{\bf{\bar C}}_u \!\buildrel \Delta \over =\! {{\bf{X}}_u}{\bf{R}}_u^{{\rm{SFT,p}}}{\bf{X}}_u^{\rm{H}} \!+\!\!\! \sum\limits_{u' \in {\mathcal{I}_u}}\!\!\! {{{\bf{X}}_{u'}}{\bf{R}}_{u'}^{{\rm{SFT,p}}}{\bf{X}}_{u'}^{\rm{H}} \!+\! \sigma _{\rm{z}}^2{{\bf{I}}_{M{N_{\rm{v}}}{N_{\rm{F}}}}}} ,
\end{equation}
and ${\mathcal{I}_u} \buildrel \Delta \over = \left\{ {u'\left| {u' \ne u,{\bf{R}}_{u'}^{\rm{B}}{\bf{R}}_u^{\rm{B}} \ne {\bf{0}}} \right.} \right\}$.

By comparing ${\bf{\bar C}}_u$ and ${\bf{ C}}$, the NMSE of each UT is only related to the statistical CSI of its own and UTs with overlapping TB domain channels along the spatial-beam dimension when only $M \to \infty $.
Therefore, when the TB domain channels of UTs $u$ and $u'$ are non-overlapping along the spatial-beam dimension, i.e., $u' \notin {\mathcal{I}_u}$, they can reuse the same pilot sequence without affecting the NMSE performance. 

We define channel overlap degree in the spatial-beam dimension between UTs $u$ and $u'$  as
\begin{equation}\label{channeloverlap-b}
{\tilde \rho _{u,u'}} \buildrel \Delta \over = \frac{{{\rm{tr}}\left\{ {{\bf{R}}_u^{{\rm{B}}}{\bf{R}}_{u'}^{{\rm{B}}}} \right\}}}{{{{\left\| {{\bf{R}}_u^{{\rm{B}}}} \right\|}_{\rm{F}}}{{\left\| {{\bf{R}}_{u'}^{{\rm{B}}}} \right\|}_{\rm{F}}}}}.
\end{equation}
It is worth noting that when we only utilize the spatial-beam dimension of TB domain statistical CSI to perform UT grouping, the computational complexity of ${\tilde \rho _{u,u'}}$ is $\mathcal{O}\left( {{N_{{\rm{an}}}}} \right)$, which is much smaller than that of ${\rho _{u,u'}}$. Moreover, UTs with overlapping TB domain channels along the spatial-beam dimension can still be allocated pilot sequences with different phase shift factors to suppress the inter-UT interference.
The corresponding UT grouping and pilot scheduling method are similar to the case of using the whole TB domain statistical CSI, in which only  ${\rho _{u,u'}}$ needs to be replaced by ${\tilde \rho _{u,u'}}$.

\subsection{Channel Prediction for Data Segment}
With the channel estimates for pilot segments, we can directly use it for DL transmit design or UL signal detection in the slow fading channel, where the channel changing over one time slot can be ignored. On the other hand, in the fast fading channel, it is not appropriate to directly apply the channel at pilot segments to  data segments \cite{6608213}. 

If the channel varies symbol by symbol, we can try to predict the channel at the data segment using the relationship between the SFT domain channel and the TB domain channel. Specifically, according to (\ref{equ-chan-sft}), the estimate of the SFT domain channel of the ${N_{\rm{F}}}$ timeslots for all UTs can be given by
\begin{equation}\label{equ-CHPre-1}
{{\bf{\hat h}}^{{\rm{SFT}}}} = \left( {{{\bf{I}}_U} \otimes {\bf{P}}} \right){{\bf{\hat h}}^{{\rm{TB}}}},
\end{equation}
where ${{\bf{\hat h}}^{{\rm{SFT}}}}\!\! \buildrel \Delta \over = \!\!{\left[ {{{\left(\! {{\bf{\hat h}}_0^{{\rm{SFT}}}} \!\right)}^{\!\rm{T}}}\!\!,\!{{\left(\! {{\bf{\hat h}}_1^{{\rm{SFT}}}} \!\right)}^{\!\rm{T}}}\!\!, \!\cdots \!,\!{{\left(\! {{\bf{\hat h}}_{U - 1}^{{\rm{SFT}}}} \!\right)}^{\!\rm{T}}}} \right]^{\!\rm{T}}} \!\!\!\!\!\in\!\! {\mathbb{C}^{M{N_{\rm{v}}}{N}U \!\times\! 1}}$, and ${\bf{\hat h}}_u^{{\rm{SFT}}} \in {\mathbb{C}^{M{N_{\rm{v}}}{N} \times 1}}$ is the estimate of the SFT domain channel of the whole frame for UT $u$. Note that the channel of the data segment to be predicted is in the current timeslot and has been contained in ${{\bf{\hat h}}^{{\rm{SFT}}}}$. Hence, the predicted space-frequency  domain channel corresponding to the ${{n_{\rm{S}}}}$-th OFDM symbol of UT $u$ in the current timeslot is given by
\begin{align}\label{equ-CHPre-2}
{\bf{\hat h}}_{u,{n_{\rm{S}}}}^{{\rm{SF}}} = {\left[ {{\bf{\hat h}}_u^{{\rm{SFT}}}} \right]_{\left( {\left( {{N_{\rm{F}}} - 1} \right){N_{\rm{S}}} + {n_{\rm{S}}}} \right)M{N_{\rm{v}}}:\hfill\atop\left( {\left( {{N_{\rm{F}}} - 1} \right){N_{\rm{S}}} + {n_{\rm{S}}} + 1} \right)M{N_{\rm{v}}} - 1}} \in {\mathbb{C}^{M{N_{\rm{v}}} \times 1}},
\end{align}
where ${n_{\rm{S}}} = 0,1, \cdots ,{N_{\rm{S}}} - 1$.

In summary, to acquire the CSI of the current timeslot, we first combine the current timeslot and the previous timeslots into a complete frame and utilize the received signal at  pilot segments to obtain the channel estimation result ${{\bf{\hat h}}^{{\rm{SFT,p}}}}$ and ${{\bf{\hat h}}^{{\rm{TB}}}}$. Then we use the estimated TB domain channel, ${{\bf{\hat h}}^{{\rm{TB}}}}$, to predict the channel at the data segment according to (\ref{equ-CHPre-1}) and (\ref{equ-CHPre-2}).

\section{CBFEM Based Channel Estimation}
To reduce the complexity, we formulate the channel estimation as a statistical inference problem and develop a CBFEM based channel estimation algorithm. Then we exploit the structure of the TB matrix and the CZT to further reduce its computational complexity.

\subsection{Algorithm}
Although the MMSE estimation in (\ref{equ-MMSE}) can achieve the optimal channel estimation performance, its complexity is unaffordable in practical systems.
	Since the statistical CSI, ${{{\bf{R}}^{{\rm{TB}}}}}$, is available,  (\ref{equ-cs}) can be rewritten as ${\bf{y}}{\rm{ = }}{\bf{\bar A}}{{{\bf{\bar h}}}^{{\rm{TB}}}}{\rm{ + }}{\bf{z}}$, where ${{{\bf{\bar h}}}^{{\rm{TB}}}}\! \in\! {\mathbb{C}^{N_{{\rm{ave}}}^{{\rm{TB}}}U \times 1}}$ consists of non-zero elements in ${{{\bf{ h}}}^{{\rm{TB}}}}$, ${\bf{\bar A}} \!\in\! {\mathbb{C}^{{M{N_{\rm{v}}}{N_{\rm{F}}} \times N_{{\rm{ave}}}^{{\rm{TB}}}U}}}$ consists of the columns of ${{\bf{ A}}}$ corresponding to the non-zero positions, and  ${N_{{\rm{ave}}}^{{\rm{TB}}}}$ is the average number of non-zero elements of ${\bf{h}}_u^{{\rm{TB}}}$ for each UT. Thus the complexity of (\ref{equ-MMSE}) is $\mathcal{O}\left( {{{\left( {M{N_{\rm{v}}}{N_{\rm{F}}}} \right)}^3} + {{\left( {M{N_{\rm{v}}}{N_{\rm{F}}}} \right)}^2}N_{{\rm{ave}}}^{{\rm{TB}}}U} \right)$. 
	Moreover, with the matrix inversion lemma, the MMSE estimate of ${{{\bf{\bar h}}}^{{\rm{TB}}}}$ can be rewritten as ${\bf{\bar h}}_{{\rm{MMSE}}}^{{\rm{TB}}} = {{\bf{\bar R}}^{{\rm{TB}}}}{\left( {{{{\bf{\bar A}}}^{\rm{H}}}{\bf{\bar A}}{{{\bf{\bar R}}}^{{\rm{TB}}}} + \sigma _{\rm{z}}^2{\bf{I}}} \right)^{ - 1}}{{\bf{\bar A}}^{\rm{H}}}{\bf{y}}$,
	where ${{\bf{\bar R}}^{{\rm{TB}}}}$	is a diagonal matrix consisting of the non-zero diagonal elements of ${{\bf{ R}}^{{\rm{TB}}}}$.
	In this case, the complexity becomes $\mathcal{O}\left(\! {{{\left( {N_{{\rm{ave}}}^{{\rm{TB}}}U} \right)}^3} \!\!+\!\! {{\left( {N_{{\rm{ave}}}^{{\rm{TB}}}U} \right)}^2}\!M{N_{\rm{v}}}{N_{\rm{F}}}} \right)$. 
	Therefore, the complexity of MMSE estimation is $\mathcal{O}\Big( {\rm{min}}\Big\{ {{\left( {M{N_{\rm{v}}}{N_{\rm{F}}}} \right)}^3} +{{\left( {M{N_{\rm{v}}}{N_{\rm{F}}}} \right)}^2} N_{{\rm{ave}}}^{{\rm{TB}}}U, {{\left( {N_{{\rm{ave}}}^{{\rm{TB}}}U} \right)}^3}\!\! +\!\! {{\left( {N_{{\rm{ave}}}^{{\rm{TB}}}U} \right)}^2}M{N_{\rm{v}}}{N_{\rm{F}}} \Big\} \Big)$.
	Although the TB domain channel exhibit sparsity in HF skywave massive MIMO-OFDM communications, when the number of UTs $U$ is relatively large, the large ${N_{{\rm{ave}}}^{{\rm{TB}}}U}$ will make the complexity of MMSE estimation still unbearable in practical systems. Therefore, we develop a low-complexity channel estimation algorithm by formulating the channel estimation as a statistical inference problem.

Many statistical inference methods can be used to solve this problem, such as message passing algorithms \cite{6940305, 8425578, 8171203,8723310,9351786,9439804,8328018,6033942}. A novel technique, named CBFEM, can unify different message passing algorithms into a single optimization framework, which has a clear objective function and can derive different solving algorithms through different constraints.
Therefore, we propose to use the CBFEM technique to estimate ${{\bf{h}}^{{\rm{TB}}}}$ by transforming the statistical inference problem into an optimization one and derive the CBFEM based channel estimation algorithm by solving this problem.

From (\ref{equ-cs}), we have
\begin{align}\label{equ-Bethe1}
&p\left( {{{\bf{h}}^{{\rm{TB}}}},{\bf{w}}\left| {\bf{y}} \right.} \right) \nonumber \\
&\!= \!\frac{1}{{p\left( {\bf{y}} \right)}}p\left( {{\bf{y}}\left| {\bf{w}} \right.} \right)p\left( {{\bf{w}}\left| {{{\bf{h}}^{{\rm{TB}}}}} \right.} \right)p\left( {{{\bf{h}}^{{\rm{TB}}}}} \right)\nonumber\\
&\!= \!\!\frac{1}{{p\left( {\bf{y}} \right)}}\!\!\!\!\prod\limits_{i = 0}^{M{N_{\rm{v}}}{N_{\rm{F}}} - 1} \!\!\!\!\!{p\left( {{y_i}\left| {{w_i}} \right.} \right)}\!\!\!\! \prod\limits_{i = 0}^{M{N_{\rm{v}}}{N_{\rm{F}}} - 1}\!\!\!\!\!\! p\left( {{w_i}\left| {{{\bf{h}}^{{\rm{TB}}}}} \right.} \right)\!\!\!\!\!\!\!	\prod\limits_{j = 0}^{{N_{\rm{an}}}{N_{\rm{de}}}{N_{\rm{do}}}U - 1}\!\!\!\!\!\!\! {p\left( {h_j^{{\rm{TB}}}} \right)} ,
\end{align}
where ${\bf{w}} \buildrel \Delta \over = {\bf{A}}{{\bf{h}}^{{\rm{TB}}}} \in {\mathbb{C}^{M{N_{\rm{v}}}{N_{\rm{F}}} \times 1}}$ is the auxiliary vector, 
$p\left( {{y_i}\left| {{w_i}} \right.} \right) = \mathcal{CN}\left( {{y_i};{w_i},\sigma _{\rm{z}}^2} \right)$,
$p\left( {{w_i}\left| {{{\bf{h}}^{{\rm{TB}}}}} \right.} \right) = \delta \left( {{w_i} - {{\bf{a}}_i}{{\bf{h}}^{{\rm{TB}}}}} \right)$, 
$p\left( {h_j^{{\rm{TB}}}} \right) = \mathcal{CN}\left( {h_j^{{\rm{TB}}};0,{{\left[ {{{\bf{R}}^{{\rm{TB}}}}} \right]}_{j,j}}} \right)$,
${{{\bf{a}}_i}}$ is the $i$-th row of ${\bf{A}}$, ${h_j^{{\rm{TB}}}}$ is the $j$-th element of ${{{\bf{h}}^{{\rm{TB}}}}}$, and ${{y_i}}$ and ${{w_i}}$ are the $i$-th element of ${\bf{y}}$ and ${\bf{w}}$, respectively. 

Variational Bayesian inference \cite{4644060,frey1998graphical} can be utilized to find a trial belief ${b\left( {{{\bf{h}}^{{\rm{TB}}}},{\bf{w}}} \right)}$ to approximate \emph{a posterior} probability density ${p\left( {{{\bf{h}}^{{\rm{TB}}}},{\bf{w}}\left| {\bf{y}} \right.} \right)}$ from a specific probability density family $\mathcal{Q}$ by minimizing the variational free energy ${F_{\rm{V}}}\left( b \right)$, i.e., $b\left( {{{\bf{h}}^{{\rm{TB}}}},{\bf{w}}} \right) =\!\!\! \mathop {\arg \min }\limits_{b\left( {{{\bf{h}}^{{\rm{TB}}}},{\bf{w}}} \right) \in \mathcal{Q}} {F_{\rm{V}}}\left( b \right)$, and ${F_{\rm{V}}}\left( b \right)$ is defined as 
\begin{align}\label{equ-Bethe-v}
{F_{\rm{V}}}\left( b \right) = {\mathbb{D}}\left\{ {b\left( {{{\bf{h}}^{{\rm{TB}}}},{\bf{w}}} \right)\left\| {p\left( {{{\bf{h}}^{{\rm{TB}}}},{\bf{w}}\left| {\bf{y}} \right.} \right)} \right.} \right\} - \ln p\left( {\bf{y}} \right),
\end{align}
where $- \ln p\left( {\bf{y}} \right)$ is termed Helmholtz free energy.
We use the Bethe approximation to limit the range of probability density family $\mathcal{Q}$ by introducing factor beliefs and variable beliefs \cite{1459044}. Denote ${b_{y,i}}\left( {{w_i}} \right)$, ${b_{w,i}}\left( {{w_i},{{\bf{h}}^{{\rm{TB}}}}} \right)$, and ${b_{h,j}}\left( {h_j^{{\rm{TB}}}} \right)$ as the factor beliefs of ${p\left( {{y_i}\left| {{w_i}} \right.} \right)}$, ${p\left( {{w_i}\left| {{{\bf{h}}^{{\rm{TB}}}}} \right.} \right)}$, and ${p\left( {h_j^{{\rm{TB}}}} \right)}$, respectively. Let ${q_{w,i}}\!\left( {{w_i}} \right)$ and ${q_{h,j}}\left( {h_j^{{\rm{TB}}}} \right)$ denote the variable beliefs of ${{w_i}}$ and ${h_j^{{\rm{TB}}}}$, respectively.
According to the Bethe approximation, we have
\begin{align}\label{equ-Bethe2}
&b\left( {{{\bf{h}}^{{\rm{TB}}}},{\bf{w}}} \right) \nonumber \\
&\!= \!\!\frac{{\prod\limits_{i = 0}^{M{N_{\rm{v}}}{N_{\rm{F}}} - 1}\!\!\!\! {{b_{y,i}}\left( {{w_i}} \right){b_{w,i}}\left( {{w_i},{{\bf{h}}^{{\rm{TB}}}}} \right)} \prod\limits_{j = 0}^{{N_{\rm{an}}}{N_{\rm{de}}}{N_{\rm{do}}}U - 1} \!\!\!\!\!\!\!\!{{b_{h,j}}\left( {h_j^{{\rm{TB}}}} \right)} }}{{\prod\limits_{i = 0}^{M{N_{\rm{v}}}{N_{\rm{F}}} - 1}\!\!\!\! {{q_{w,i}}\left( {{w_i}} \right)\prod\limits_{j = 0}^{{N_{\rm{an}}}{N_{\rm{de}}}{N_{\rm{do}}}U - 1}\!\!\!\! {{{\left( {{q_{h,j}}\left( {h_j^{{\rm{TB}}}} \right)} \right)}^{M{N_{\rm{v}}}{N_{\rm{F}}}}}} } }}, \nonumber \\
& \qquad\qquad\qquad\qquad\qquad\qquad\qquad\forall b\left( {{{\bf{h}}^{{\rm{TB}}}},{\bf{w}}} \right) \in \mathcal{Q}.
\end{align}
Moreover, the Bethe approximation needs to fulfill the marginalization consistency constraints
\begin{subequations}\label{equ-Behte3}
	\begin{equation}
	{{b_{y,i}}\left( {{w_i}} \right)}  = \int { {{b_{w,i}}\left( {{w_i},{{\bf{h}}^{{\rm{TB}}}}} \right)}}d{\bf{h}}^{{\rm{TB}}}  = {q_{w,i}}\left( {{w_i}} \right),
	\end{equation}
	\begin{equation}
	{{b_{h,j}}\!\left( {h_j^{{\rm{TB}}}} \right)}\! = \!\!\!\int {\!\! {{b_{w,i}}\!\left( {{w_i},{{\bf{h}}^{{\rm{TB}}}}} \right)}  d{{w_i}}d{\bf{h}}_{\backslash h_j^{{\rm{TB}}}}^{{\rm{TB}}}}  \!=\! {q_{h,j}}\!\left( {h_j^{{\rm{TB}}}} \right),
	\end{equation}
\end{subequations}
where the normalization and non-negative constraints are omitted for brevity since they hold for any valid belief. According to \cite{2002Expectation}, we can relax the constraints in (\ref{equ-Behte3}) into the first-order and 
second-order moment consistency constraints as follows
\begin{subequations}\label{equ-Bethe4}
	\begin{equation}\label{equ-Bethe4-1}
		\mathbb{E}\left\{ {{w_i}\left| {{b_{y,i}}} \right.} \right\} = \mathbb{E}\left\{ {{w_i}\left| {{b_{w,i}}} \right.} \right\} = \mathbb{E}\left\{ {{w_i}\left| {{q_{w,i}}} \right.} \right\}
	\end{equation}
	\begin{equation}\label{equ-Bethe4-2}
		\mathbb{E}\left\{ {h_j^{{\rm{TB}}}\left| {{b_{h,j}}} \right.} \right\} = \mathbb{E}\left\{ {h_j^{{\rm{TB}}}\left| {{b_{w,i}}} \right.} \right\} = \mathbb{E}\left\{ {h_j^{{\rm{TB}}}\left| {{q_{h,j}}} \right.} \right\}
	\end{equation}
	\begin{equation}\label{equ-Bethe4-3}
		\mathbb{E}\left\{ {{{\left| {{w_i}} \right|}^2}\left| {{b_{y,i}}} \right.} \right\} = \mathbb{E}\left\{ {{{\left| {{w_i}} \right|}^2}\left| {{b_{w,i}}} \right.} \right\} = \mathbb{E}\left\{ {{{\left| {{w_i}} \right|}^2}\left| {{q_{w,i}}} \right.} \right\}
	\end{equation}
	\begin{equation}\label{equ-Bethe4-4}
		\mathbb{E}\left\{\! {{{\left| {h_j^{{\rm{TB}}}} \right|}^2}\left| {{b_{h,j}}} \right.}\!\! \right\}\! = \!\frac{{\sum\limits_{i = 0}^{M{N_{\rm{v}}}{N_{\rm{F}}} - 1}\!\!\!\!\!\! {\mathbb{E}\left\{\! {{{\left| {h_j^{{\rm{TB}}}} \right|}^2}\!\left| {{b_{w,i}}} \right.} \!\right\}} }}{{M{N_{\rm{v}}}{N_{\rm{F}}}}} \!\! = \!\!\mathbb{E}\left\{\! {{{\left| {h_j^{{\rm{TB}}}} \right|}^2}\!\left| {{q_{h,j}}} \right.} \!\right\}
	\end{equation}
\end{subequations}
where (\ref{equ-Bethe4-4}) is an approximate second-order moment consistency constraint due to the fact that the belief ${{b_{w,i}}}$ contains the variable ${h_j^{{\rm{TB}}}}$ for any $i$.

Substituting (\ref{equ-Bethe2}) into (\ref{equ-Bethe-v}), we can obtain the Bethe free energy as follows
\begin{align}
	&{F_{\rm{B}}}\left( b \right) =\!\!\!\!\!\! \sum\limits_{i = 0}^{M{N_{\rm{v}}}{N_{\rm{F}}} - 1}\!\!\!\!\!\!\! \!{{\mathbb{D}}\left\{ {{b_{y,i}}\left\| {p\left( {{y_i}\!\left| {{w_i}} \right.} \right)} \right.} \!\right\}}\!  + \!\!\!\!\!\!\!\!\sum\limits_{i = 0}^{M{N_{\rm{v}}}{N_{\rm{F}}} - 1} \!\!\!\!\!\!\!\!{{\mathbb{D}}\left\{ {{b_{w,i}}\left\| {p\left( {{w_i}\!\left| {{{\bf{h}}^{{\rm{TB}}}}} \right.} \right)} \right.} \!\!\right\}} \nonumber\\
	&+ \!\!\!\!\!\!\sum\limits_{j = 0}^{{N_{\rm{an}}}{N_{\rm{de}}}{N_{\rm{do}}}U - 1} \!\!\!\!\!\!\!\!\!\!\!\!\!{{\mathbb{D}}\left\{\! {{b_{h,j}}\!\left\| {p\left( {h_j^{{\rm{TB}}}} \right)} \right.} \!\right\}} \!\! +\! {\mathbb{H}}\left\{ {{q_{w,i}}} \right\}\! +\!\!\!\!\! \!\!\!\!\!\!\!\!\!\!\sum\limits_{j = 0}^{{N_{\rm{an}}}{N_{\rm{de}}}{N_{\rm{do}}}U - 1} \!\!\!\!\!\!\!\!\!\!\!\!\!{M{N_{\rm{v}}}{N_{\rm{F}}}{\mathbb{H}}\left\{\! {{q_{h,j}}}\! \right\}} ,
\end{align}
Finally, we transform the channel estimation into the following CBFEM question
\begin{equation}\label{equ-Bethe5}
\mathop {\min }\limits_{\scriptstyle\left\{ {{b_{y,i}}} \right\},\left\{ {{b_{w,i}}} \right\},\{ {{b_{h,j}}}\},\atop
	\scriptstyle\left\{ {{q_{w,i}}} \right\},\{ {{q_{h,j}}} \}} {F_{\rm{B}}}\left( b \right) \quad {\rm{s}}{\rm{.t}}{\rm{.}}(\ref{equ-Bethe4}).
\end{equation} 
We can utilize Lagrange multipliers to solve the CBFEM question, which is given in Appendix C. The resulting CBFEM based channel estimation algorithm is summarized in Algorithm 2, where  ${b_h}  \buildrel \Delta \over =  \prod\nolimits_{j = 0}^{{N_{{\rm{an}}}}{N_{{\rm{de}}}}{N_{{\rm{do}}}}U - 1} {{b_{h,j}}} $, ${\bf{e}} \buildrel \Delta \over =  {\left[ {1,1, \cdots ,1} \right]^{\rm{T}}} \in {\mathbb{C}^{{N_{{\rm{an}}}}{N_{{\rm{de}}}}{N_{{\rm{do}}}}U \times 1}}$ and ${\left(  \cdot  \right)^{ \circ  - 1}}$ denotes the element-wise inverse. Moreover, during the iterations, the damping factor can be leveraged to ensure the convergence of the algorithm \cite{9351786}.

\IncMargin{0.2em}
{
\begin{algorithm}
	\SetAlgoNoLine 
    \SetNlSty{textbf}{}{}
    \SetAlgoNlRelativeSize{-2}
	\caption{CBFEM based channel estimation algorithm}
	
	\KwIn{
		$\bf{A}$, $\bf{y}$, ${\sigma _{\rm{p}}^2}$, ${\sigma _{\rm{z}}^2} $, $p\left( {{{\bf{h}}^{{\rm{TB}}}}} \right) = \mathcal{CN}\left( {{{\bf{h}}^{{\rm{TB}}}};{\bf{0}},{{\bf{R}}^{{\rm{TB}}}}} \right)$ 
	}
	
	{\bfseries Initialize:}  ${b_h} = p\left( {{{\bf{h}}^{{\rm{TB}}}}} \right), {{\bm{\eta }}^{h,{b_h}}} = {\bf{0}}$ 
	
	\Repeat{\rm{the termination condition is fulfilled}}{
		${{\bm{\eta }}^{h,{b_w}}} =  - {\rm{Var}}{\left\{ {{{\bf{h}}^{{\rm{TB}}}}\left| {{b_h}} \right.} \right\}^{ \circ  - 1}} - \frac{1}{{M{N_{\rm{v}}}{N_{\rm{F}}}}}{{\bm{\eta }}^{h,{b_h}}}$ \\
		${{\bm{\eta }}^{h,{b_h}}} = M{N_{\rm{v}}}{N_{\rm{F}}}\!{\left( \!{\left(\! {\sigma _{\rm{p}}^2{{\bf{e}}^{\rm{T}}}\!{{\left( {{{\bm{\eta }}^{h,{b_w}}}}\! \right)}^{ \!\circ  - \!1}}\!\!\!\!\! -\! \sigma _{\rm{z}}^2} \right)\sigma _{\rm{p}}^{ - 2}{\bf{e}}\! -\! {{\left( {{{\bm{\eta }}^{h,{b_w}}}} \!\right)}^{ \!\circ  - \!1}}} \!\right)^{\! \!\circ  - \!1}}$ \\
		${{\bm{\tilde \tau }}^{h,{b_w}}} = \mathbb{E}\left\{ {{{\bf{h}}^{{\rm{TB}}}}\left| {{b_h}} \right.} \right\} \circ {\rm{Var}}{\left\{ {{{\bf{h}}^{{\rm{TB}}}}\left| {{b_h}} \right.} \right\}^{ \circ  - 1}}$ \\
		${\bm{\kappa }} = {{\bm{\tilde \tau }}^{h,{b_w}}} \circ {\left( {{{\bm{\eta }}^{h,{b_w}}}} \right)^{ \circ  - 1}}$\\
		${\bm{\psi }} = {\bf{A}\bm{\kappa }}$ \\
		${\bm{\varpi }} = \frac{1}{{M{N_{\rm{v}}}{N_{\rm{F}}}\sigma _{\rm{p}}^2}}{{\bf{A}}^{\rm{H}}}\left( {{\bf{y}} + {\bm{\psi }}} \right) - {\bm{\kappa }} $ \\	
		${b_h} \!\propto\! p\left( {{{\bf{h}}^{{\rm{TB}}}}} \right){\cal C}{\cal N}\left( {{{\bf{h}}^{{\rm{TB}}}};{\bm{\varpi }},{\rm{diag}}\left\{ { - {{\left( {{{\bm{\eta }}^{h,{b_h}}}} \right)}^{ \circ  - 1}}} \!\right\}} \right)$ \\
		
	}

	\KwOut{ ${{\bf{\hat h}}^{{\rm{TB}}}} = \mathbb{E}\left\{ {{{\bf{h}}^{{\rm{TB}}}}\left| {{b_h}} \right.} \right\}$}
\end{algorithm}
}

\subsection{Low-Complexity Implementation}
The computational complexity of each iteration of the proposed CBFEM based channel estimation algorithm mainly comes from steps 7 and 8. In this subsection, we focus on these two steps and further reduce their complexity.

Recalling (\ref{equ-stru-P}) and (\ref{equ-stru-A}), steps 7 and 8 of the CBFEM based channel estimation algorithm can be rewritten as
\begin{align}\label{equ-L1}
{\bm{\psi }} = {\sigma _{\rm{p}}}{{\bf{X}}_{\rm{c}}}\left( {{\bf{\tilde D}} \otimes \left( {{\bf{V}}\left( {{\bf{\bar U}} \otimes {{\bf{I}}_{{N_{{\rm{an}}}}}}} \right)} \right)} \right)\left[ {{{\bf{S}}_0}, \cdots ,{{\bf{S}}_{U - 1}}} \right]{\bm{\kappa }},
\end{align}
\begin{align}\label{equ-L2}
{\bm{\varpi}}  &\!= \!\frac{1}{{M{N_{\rm{v}}}{N_{\rm{F}}}{\sigma _{\rm{p}}}}}{\left[ {{{\bf{S}}_0}, \cdots\! ,{{\bf{S}}_{U - 1}}} \right]^{\rm{H}}}\!\!\left( \!{{{{\bf{\tilde D}}}^{\rm{H}}}\! \otimes\! \left(\! {\left( {{{{\bf{\bar U}}}^{\rm{H}}} \!\otimes\! {{\bf{I}}_{{N_{{\rm{an}}}}}}} \right){{\bf{V}}^{\rm{H}}}} \right)} \!\right) \nonumber \\
&\qquad\qquad\qquad\qquad\qquad\qquad\quad\times {\bf{X}}_{\rm{c}}^{\rm{H}}\left( {{\bf{y}} + {\bm{\psi }}} \right) - {\bm{\kappa }}.
\end{align}
Since ${{\bf{V}}_{k}}$ and ${\bf{\tilde D}}$ are not DFT matrices, fast Fourier transform (FFT) cannot be directly used. In this case, the CZT \cite{1162034} can be utilized to reduce the computational complexity of (\ref{equ-L1}) and (\ref{equ-L2}). Specifically, ${{\bf{V}}_{k}}$ and ${\bf{\tilde D}}$ can be rewritten as
\begin{subequations}\label{equ-L3}
	\begin{align}
		&\!\!\!{{\bf{V}}_k} = {\rm{diag}}\left\{ {{{\bm{\xi }}_{{W_{{\rm{S}},k}},{N_{{\rm{an}}}},0,M}}} \right\}{\bf{F}}_{{N_{\left( {\rm{S}} \right)}} \times M}^{\rm{H}} \nonumber \\
		&\!\!\!\times\!{\rm{diag}}\!\!\left\{ \!{{{{\bm{\tilde \xi }}}_{{W_{{\rm{S}},k}},M,{N_{{\rm{an}}}},{N_{\left( {\rm{S}}  \right)}}}}\!} \!\right\}\!{{\bf{F}}_{{N_{\left( {\rm{S}} \right)}} \!\times\! {N_{{\rm{an}}}}}}\!{\rm{diag}}\!\left\{\! {{{\bm{\xi }}_{{W_{{\rm{S}},k}},0,0,{N_{{\rm{an}}}}}}} \!\!\right\}\!\!,
	\end{align}
	\begin{align}
		&\!\!\!{\bf{\tilde D}} ={\rm{diag}}\left\{ {{{\bm{\xi }}_{{W_{\rm{T}}},  {N_{{\rm{do}}}}, \frac{{{n_{\rm{p}}}{N_{{\rm{do}}}}}}{{{N_{\rm{S}}}}},{N_{\rm{F}}}}}} \right\}{\bf{F}}_{{N_{\left( {\rm{T}} \right)}} \times {N_{\rm{F}}}}^{\rm{H}} \nonumber \\
		&\!\!\!\! \times\!{\rm{diag}}\!\! \left\{\! {{{{\bm{\tilde \xi }}}_{\!{W_{\!\rm{T}}},{N_{\!\rm{F}}},{N_{{\rm{do}}}},{N_{\!\left(  {\rm{T}}  \right)}}}}} \!\!\right\}\!\! {{\bf{F}}_{{N_{\!\left(  {\rm{T}}  \right)}} \!\times\! {N_{{\rm{do}}}}}}\!{\rm{diag}}\! \left\{\! {{{\bm{\xi }}_{\!{W_{\!\rm{T}}},-\!{\frac{{2{n_{\rm{p}}}}}{{{N_{\rm{S}}}}}},0,{N_{{\rm{do}}}}}}}\!\! \right\}\!\!,
	\end{align}
\end{subequations}
respectively, where ${{\bf{F}}_{N}}$ denotes the ${N}$-dimensional unitary DFT matrix, ${{\bf{F}}_{N \times G}}$ denotes the matrix composed of the first $G$ ($ \le N$) columns of ${{\bf{F}}_N}$,
${N_{\left(\rm{S}\right)}}$ and ${N_{\left(\rm{T}\right)}}$ are integers greater than or equal to $M + { N_{{\rm{an}}}} - 1$ and  ${ N_{{\rm{F}}}} + { N_{{\rm{do}}}} - 1$, respectively.
${W_{{\rm{S,}}k}} \buildrel \Delta \over = {e^{ - {\bar \jmath}\frac{{4\pi \left( {{f_{\rm{c}}} + k\Delta f} \right)\Delta \tau }}{{{N_{{\rm{an}}}}}}}}$, ${W_{\rm{F}}} \buildrel \Delta \over = {e^{ - {\bar \jmath}\frac{{2\pi {N_{\tau}}}}{{{N_{{\rm{de}}}}{N_{\rm{v}}}}}}}$, ${W_{\rm{T}}} \buildrel \Delta \over = {e^{{\bar \jmath}\frac{{2\pi {N_{\rm{d}}}{N_{\rm{S}}}}}{{{N_{{\rm{do}}}}N}}}}$,
${{\bm{\xi }}_{W,{\alpha _1},{\alpha _2},N}} \in {\mathbb{C}^{N \times 1}}$ with ${\left[ {{{\bm{\xi }}_{W,{\alpha _1},{\alpha _2},N}}} \right]_i} = {W^{{{\left( {{i^2} - {\alpha _1}i - {\alpha _2}} \right)} \mathord{\left/
				{\vphantom {{\left( {{i^2} - {\alpha _1}i - {\alpha _2}} \right)} 2}} \right.
				\kern-\nulldelimiterspace} 2}}}$, and
${{{\bm{\tilde \xi }}}_{W,{N_1},{N_2},N}} \buildrel \Delta \over = \sqrt N {\bf{F}}_N^{}{\left[ {{\bm{\xi }}_{W,0,0,{N_1}}^{\rm{H}},{\bf{0}},{\bm{\xi }}_{W,2\left( {{N_2} - 1} \right), - {{\left( {{N_2} - 1} \right)}^2},{N_2} - 1}^{\rm{H}}} \right]^{\rm{T}}} \in {\mathbb{C}^{N \times 1}}$.

With (\ref{equ-L1}), (\ref{equ-L2}) and (\ref{equ-L3}), steps 7 and 8 can be implemented efficiently and the complexities are  respectively reduced to $\mathcal{O}\Big( {N_{{\rm{an}}}}{N_{{\rm{do}}}}{\bar N_{{\rm{de}}}}{{\log }_2}{\bar N_{{\rm{de}}}} + {N_{\rm{v}}}{N_{{\rm{do}}}}N_{\left(\rm{S}\right)}{{\log }_2}N_{\left(\rm{S}\right)} + M{N_{\rm{v}}}N_{\left(\rm{T}\right)}{{\log }_2}N_{\left(\rm{T}\right)} \Big)$ and 
$\mathcal{O}\Big( {N_{\rm{v}}}{N_{\rm{F}}}N_{\left(\rm{S}\right)}{{\log }_2}N_{\left(\rm{S}\right)} + {N_{{\rm{an}}}}{N_{\rm{F}}}{\bar N_{{\rm{de}}}}{{\log }_2}{\bar N_{{\rm{de}}}} + {N_{{\rm{an}}}}{\bar N_{{\rm{de}}}}N_{\left(\rm{T}\right)}{{\log }_2}N_{\left(\rm{T}\right)} \Big)$, where ${\bar N_{{\rm{de}}}}  \buildrel \Delta \over = \left\lfloor {{{{N_{\rm{v}}}} \mathord{\left/
			{\vphantom {{{N_{\rm{v}}}} {{N_\tau }}}} \right.
			\kern-\nulldelimiterspace} {{N_\tau }}}} \right\rfloor {N_{{\rm{de}}}}$.

\section{Simulation Results}
In this section, we provide simulation results to illustrate the performance of the proposed channel acquisition approach for HF skywave massive MIMO-OFDM communications. The main simulation parameters are given in Table \ref{tab1}. 
We generate UTs at a distance of 2000 km from the BS, and azimuth angles of the UTs seen from the BS are generated in the interval (-90° 90°). A commercial ray-tracing software, Proplab-Pro version 3.1, is utilized to generate the realistic HF skywave channel for each UT independently\cite{proplab}. Specifically, we use the software to get parameters of each propagation path between the BS and each UT, including the path gain, the propagation delay, the azimuth and elevation AoA. The maximum Doppler comes from the ionosphere and UT mobility, i.e., ${\nu _{\max }} = \frac{\nu _{{\rm{iono}}}}{2} + \frac{{{v_u}}}{c}{f_c}$,
where ${\nu _{{\rm{iono}}}}$ is the Doppler spread imposed by the ionosphere and ${{v_u}}$ is the velocity of UT $u$. In the simulation, ${\nu _{{\rm{iono}}}}$ is set to be 0.5 Hz for moderate ionospheric conditions at mid-latitude regions \cite{ITU}, the Doppler frequency of each path is randomly generated within the range $\left[ { - {\nu _{\max }},{\nu _{\max }}} \right)$, and ${N_{{\rm{d}}}}$ is set to be 8 according to the maximum Doppler. Then using the generated path parameters and randomly generated initial phases for paths, the SFT domain channel of each UT is generated according to (\ref{equ-chan1}) and the statistical CSI ${{{\bf{R}}^{{\rm{TB}}}}}$ is acquired by the method proposed in \cite{LAA}.

Note that ${{\bf{\hat h}}^{{\rm{SFT,p}}}}$ consists of the estimated spatial-frequency domain channels of pilot segments in ${N_{\rm{F}}}$ timeslots and only the spatial-frequency domain channel of the last timeslot (i.e., the current timeslot) is needed for DL transmit design or UL signal detection. Therefore, in addition to the NMSE of all timeslots defined in (\ref{equ-nmse-ini}), we will also show the NMSE between the estimated channel and the real channel in the current timeslot. 
The signal-to-noise ratio (SNR) in the simulation refers to the received SNR, whose range is set to be [-10 20] dB. Such a range is realizable in practice since each UT only needs about 50 W of transmit power to achieve a SNR of up to 20 dB under the given simulation parameters.
We refer to the UT grouping algorithm based on the channel overlap degree (\ref{channeloverlap-bdd}) and (\ref{channeloverlap-b}) as TB-UG and B-UG, respectively, and the number of groups to divide is set to be $S = {\left\lfloor {{{{N_{\rm{v}}}} \mathord{\left/{\vphantom {{{N_{\rm{v}}}} {{N_{\tau}}}}} \right.\kern-\nulldelimiterspace} {{N_{\tau}}}}} \right\rfloor } $. The performance of random UT grouping (Random-UG) is also provided for comparison, and we refer to the CBFEM based channel estimation algorithm as the CBFEM-CE algorithm.

\begin{table}[htbp]
	\renewcommand\arraystretch{1}
	\caption{Simulation Parameters}
	\label{table}
	\centering
	\setlength{\tabcolsep}{12pt}
	\begin{tabular}{cc}
		\toprule
		Parameter & Value  \\ 
		\midrule
		Carrier frequency ${{f_{\rm{c}}}}$   & 16 MHz  \\
		Subcarrier spacing ${\Delta f }$ & 250 Hz \\
		Number of subcarriers ${{N_{\rm{c}}}}$ & 2048 \\
		Length of CP  ${{N_{\rm{g}}}}$ & 512 \\
		Number of valid subcarriers  ${{N_{\rm{v}}}}$ & 1536 \\
		Number of BS antennas $M$ &128 \\
		BS antenna spacing $d$ & 9 m \\
		Size of frame  $\left( {{N_{\rm{F}}},{N_{\rm{S}}},{n_{\rm{p}}}} \right)$ & (8, 14, 6) \\
		Number of UTs $U$ & 64 \\
		Velocity of UTs $v_u$ & 30\ /\ 100\ /\ 250 km/h \\
		\bottomrule	
	\end{tabular}
	\label{tab1}
\end{table}

\begin{figure}[htbp]
	\centering
	\subfigure[]{
		\begin{minipage}{8.89cm}
			\centering
			\includegraphics[width=0.98\textwidth]{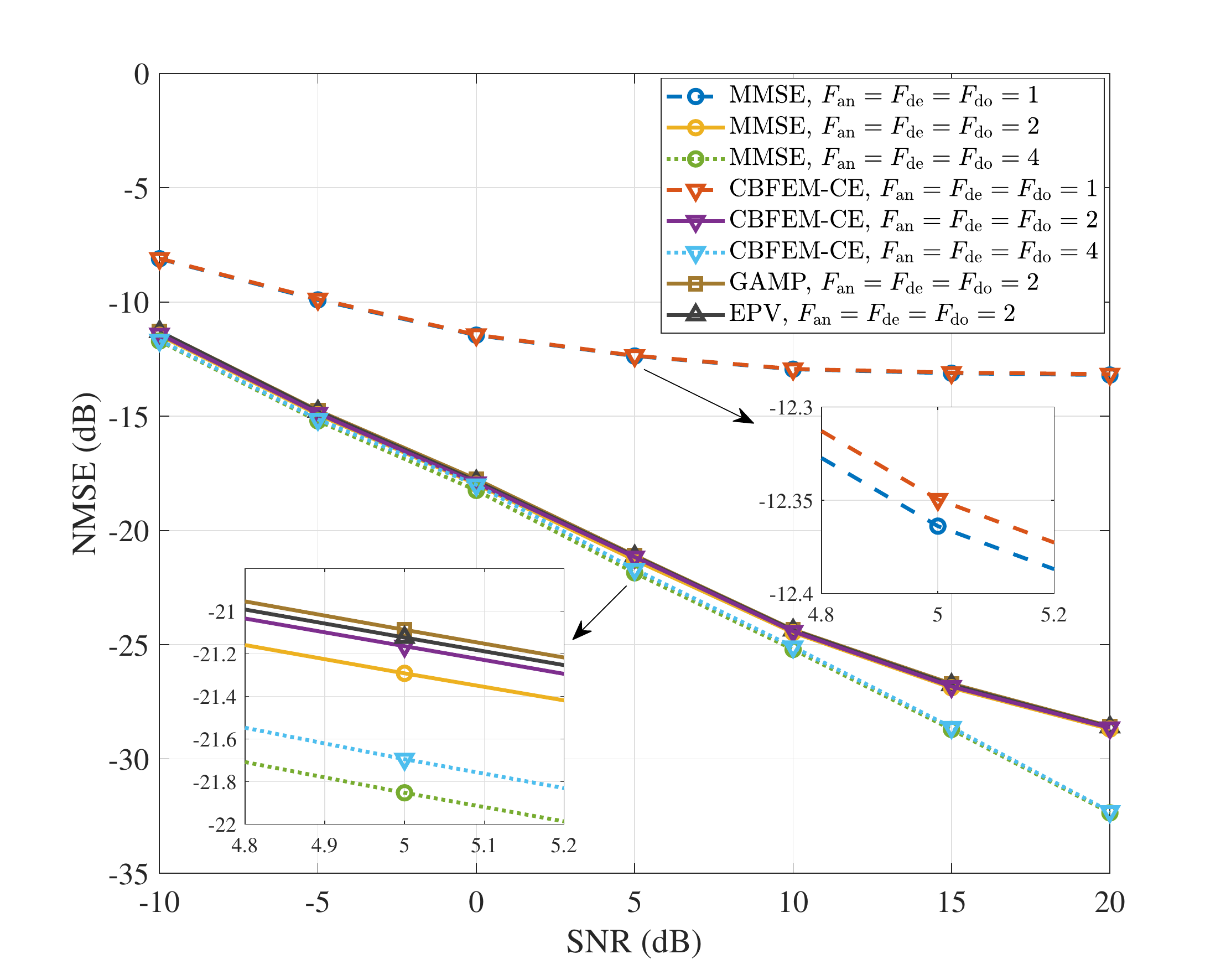}
			\label{fig-all}
		\end{minipage}
		
	}
	\subfigure[]{
		\begin{minipage}{8.89cm}
			\centering
			\includegraphics[width=0.98\linewidth]{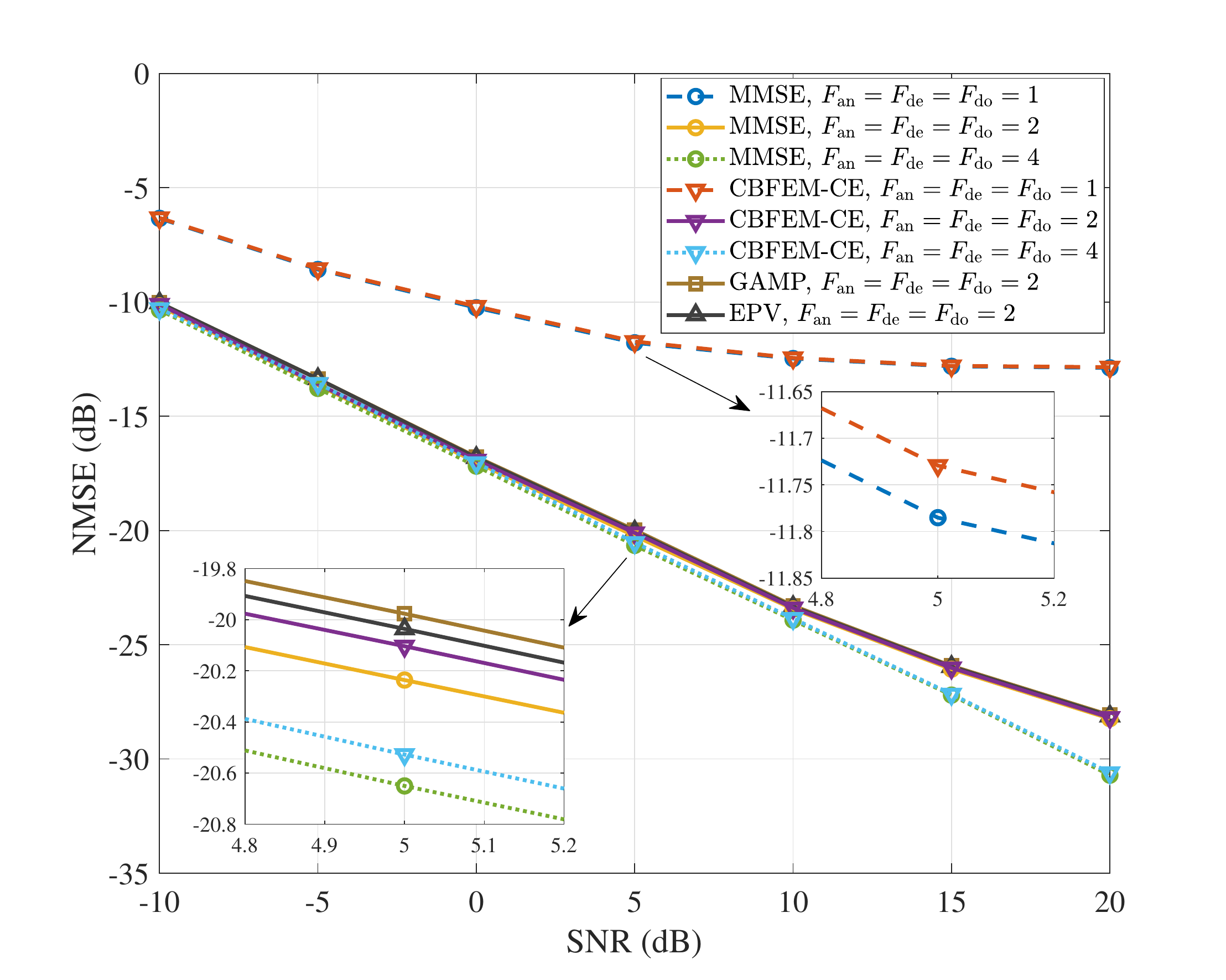}
			\label{fig-last}
		\end{minipage}
	}
	\caption{The NMSE performance of different algorithms with different fine factors. (a) NMSE of all timeslots; (b) NMSE of current timeslot.}
	\label{fig-nmse_snr}
\end{figure}

We first compare the NMSE performance of different algorithms with different fine factors. 
The MMSE estimation, the generalized approximate message passing (GAMP) \cite{6033942},  and the expectation propagation variant (EPV) \cite{9351786} algorithms are included along with the proposed CBFEM-CE algorithm.
Since the simulation parameters in Table \ref{tab1} will make the complexity of the MMSE estimation unbearable, we use a set of smaller parameters only in this simulation. Specifically, we set $M=64$, ${N_{\rm{v}}} = 128$, and the number of UTs is set to be 32. The simulation results are shown in Fig. \ref{fig-nmse_snr}, where the UT grouping algorithm is TB-UG and the UT velocity is 100 km/h.
From the figure, larger fine factors can achieve better performance due to the improved accuracy of the proposed TB based channel model.
Furthermore, the performance of CBFEM-CE, GAMP, and EPV algorithms can approach that of the optimal MMSE estimation with a very small gap. 
Moreover, the proposed CBFEM-CE algorithm can be implemented with much lower complexity by using the CZT and the structure of the TB matrix.

\begin{table}[htbp]
	\renewcommand\arraystretch{1.2}
	\caption{Computational Complexities of Different Channel Estimation Algorithms}
	\label{table-complexity}
	\centering
	\setlength{\tabcolsep}{8pt}
	{
		\begin{tabular}{c|c}
			\hline
			{\bfseries Algorithm} &  {\bfseries Complexity} \\ 
			\specialrule{0.04em}{1.2pt}{1.2pt} 
			MMSE   & 	\makecell[c]{$\mathcal{O}\Big( {\rm{min}}\Big\{ {{\left( {M{N_{\rm{v}}}{N_{\rm{F}}}} \right)}^3} + {{\left( {M{N_{\rm{v}}}{N_{\rm{F}}}} \right)}^2} N_{{\rm{ave}}}^{{\rm{TB}}}U,$ \\ ${{\left( {N_{{\rm{ave}}}^{{\rm{TB}}}U} \right)}^3} + {{\left( {N_{{\rm{ave}}}^{{\rm{TB}}}U} \right)}^2}M{N_{\rm{v}}}{N_{\rm{F}}} \Big\} \Big)$} \\
			\specialrule{0.04em}{1.2pt}{1.2pt} 
			\makecell[c]{	CBFEM-CE \\ (per iteration)}  & \makecell[c]{$\mathcal{O}\big(\left( {{N_{{\rm{an}}}}{N_{{\rm{do}}}} + {N_{{\rm{an}}}}{N_{\rm{F}}}} \right){{\bar N_{{\rm{de}}}}{{\log }_2}{\bar N_{{\rm{de}}}}}$ \\
				$	+ \left( {{N_{\rm{v}}}{N_{{\rm{do}}}} + {N_{\rm{v}}}{N_{\rm{F}}}} \right){{N_{\left( {\rm{S}} \right)}}{{\log }_2}{N_{\left( {\rm{S}} \right)}}}$ \\
				$ + \left( {M{N_{\rm{v}}} + {N_{{\rm{an}}}}{{\bar N}_{{\rm{de}}}}} \right) {{N_{\left( {\rm{T}} \right)}}{{\log }_2}{N_{\left( {\rm{T}} \right)}}}\big)$}\\ 
			\specialrule{0.04em}{1.2pt}{1.2pt} 
			\makecell[c]{	GAMP / EPV  \\(per iteration)}& $\mathcal{O}\left( {M{N_{\rm{v}}}{N_{\rm{F}}}N_{{\rm{ave}}}^{{\rm{TB}}}U} \right)$ \\
			\hline	
	\end{tabular}}
	\label{tab2}
\end{table}

\begin{figure}[htbp]
	\centering
	\includegraphics[width=0.98\linewidth]{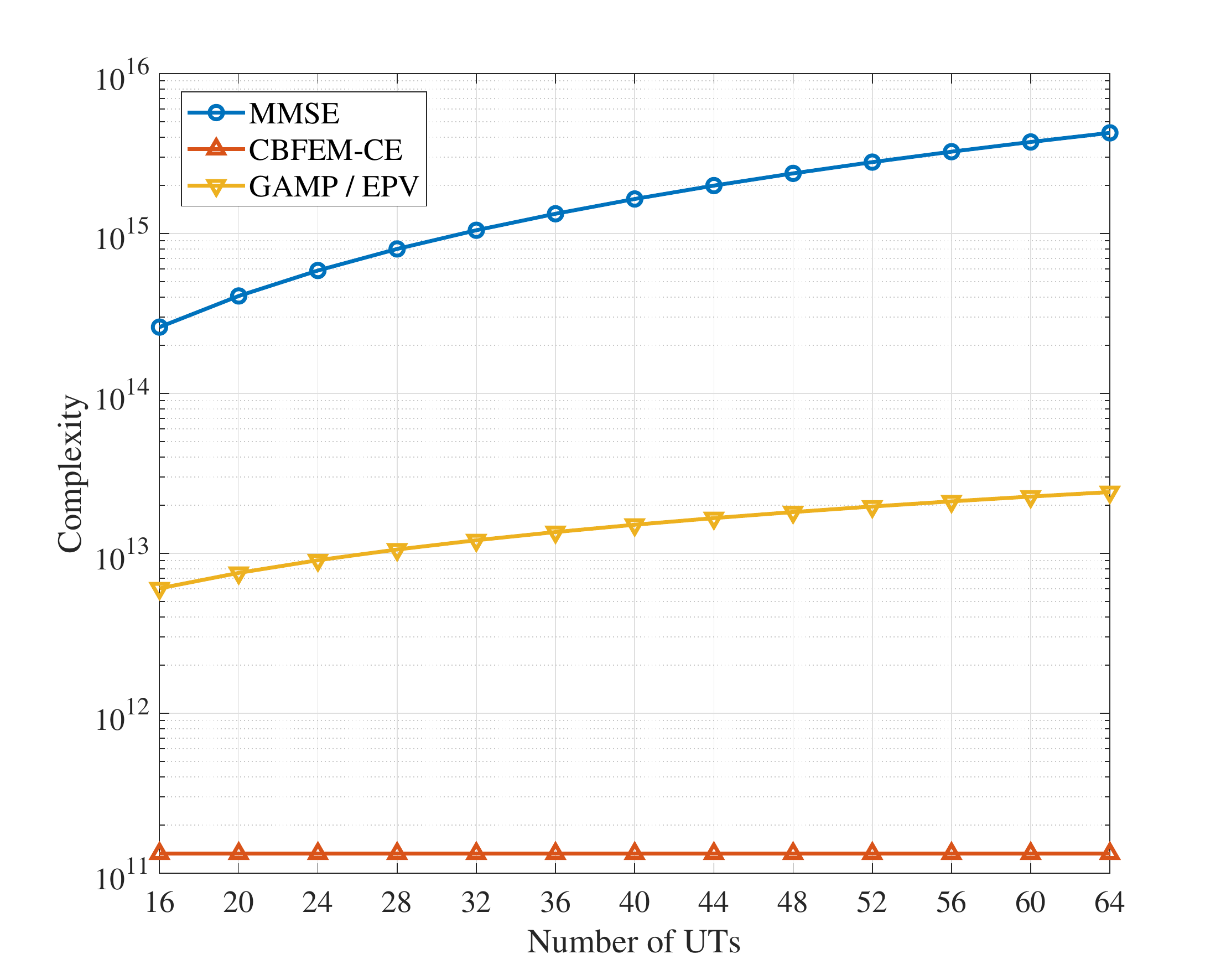}
	\caption{The complexities of different algorithms versus the number of UTs.}
	\label{fig-complexity}
\end{figure}

	The computational complexity of different channel estimation algorithms are summarized in Table \ref{tab2} and are plotted in Fig. 3 under different numbers of UTs, where ${F_{{\rm{an}}}} = {F_{{\rm{de}}}} = {F_{{\rm{do}}}} = 2$ and the number of iterations for CBFEM-CE, GAMP, and EPV algorithms are set to 300 (the number of iterations required is different for different SNRs and system configurations). From the figure, the complexity of MMSE is the highest due to the matrix-inversion. On the other hand, owing to the utilization of the structure of the TB matrix and the CZT, the complexity of CBFEM-CE algorithm is the lowest among all the algorithms.

	Fig. 4 shows the convergence performance of different algorithms  under different SNRs, where the UT velocity is 100 km/h, ${F_{{\rm{an}}}} = {F_{{\rm{de}}}} = {F_{{\rm{do}}}} = 2$, and the UT grouping algorithm is TB-UG.
	From the figure, the NMSE performance of the CBFEM-CE algorithm improves as the number of iterations increases until the convergence. Moreover, the CBFEM-CE algorithm has similar convergence behavior to that of GAMP and EPV algorithms with much lower complexity.

\begin{figure}[htbp]
	\centering
	\includegraphics[width=0.98\linewidth]{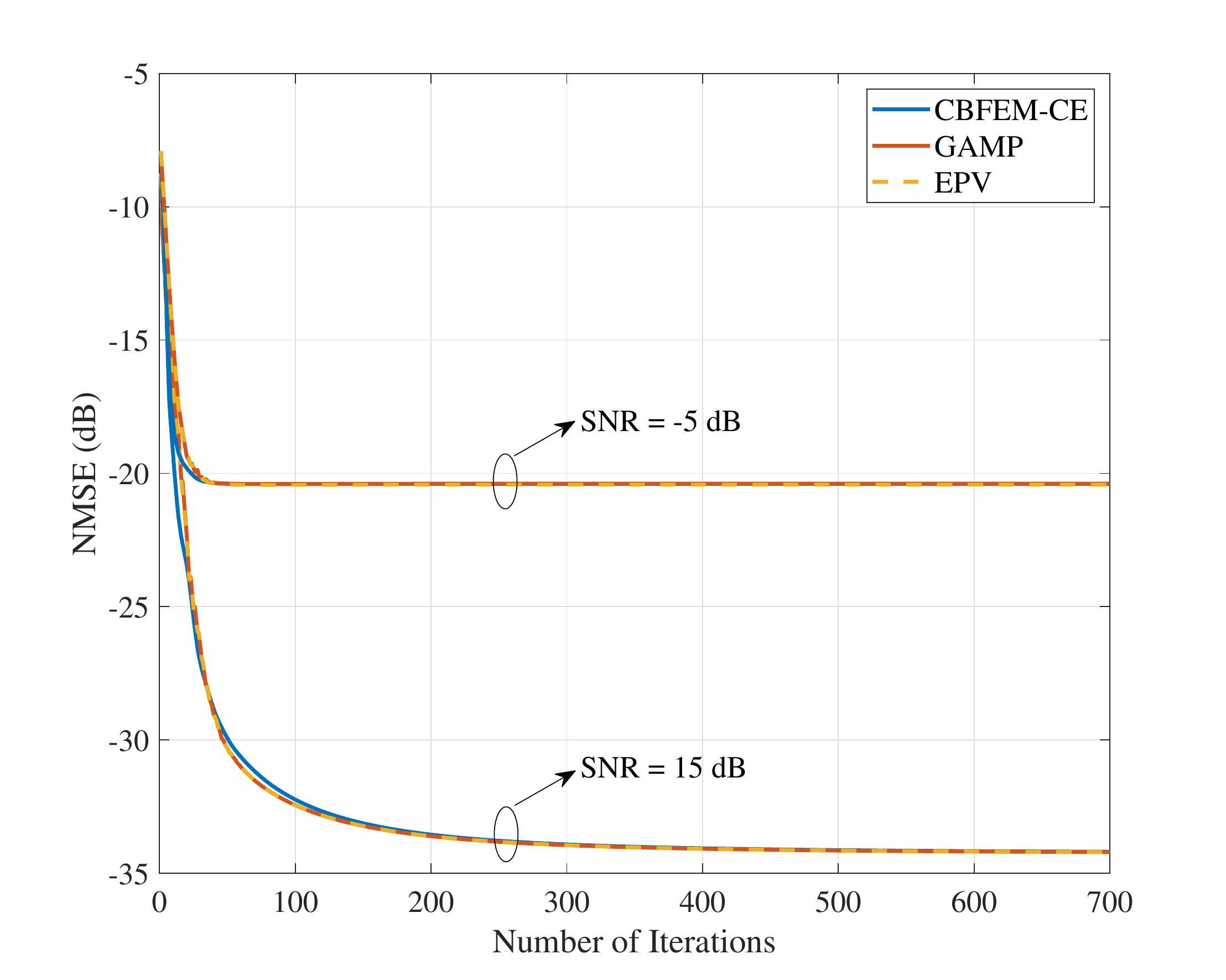}
	\caption{The convergence performance of different algorithms under different SNRs.}
	\label{fig-convergence}
\end{figure}

\begin{figure}[htbp]
	\centering
	\subfigure[]{
		\begin{minipage}{8.89cm}
			\centering
			\includegraphics[width=0.98\textwidth]{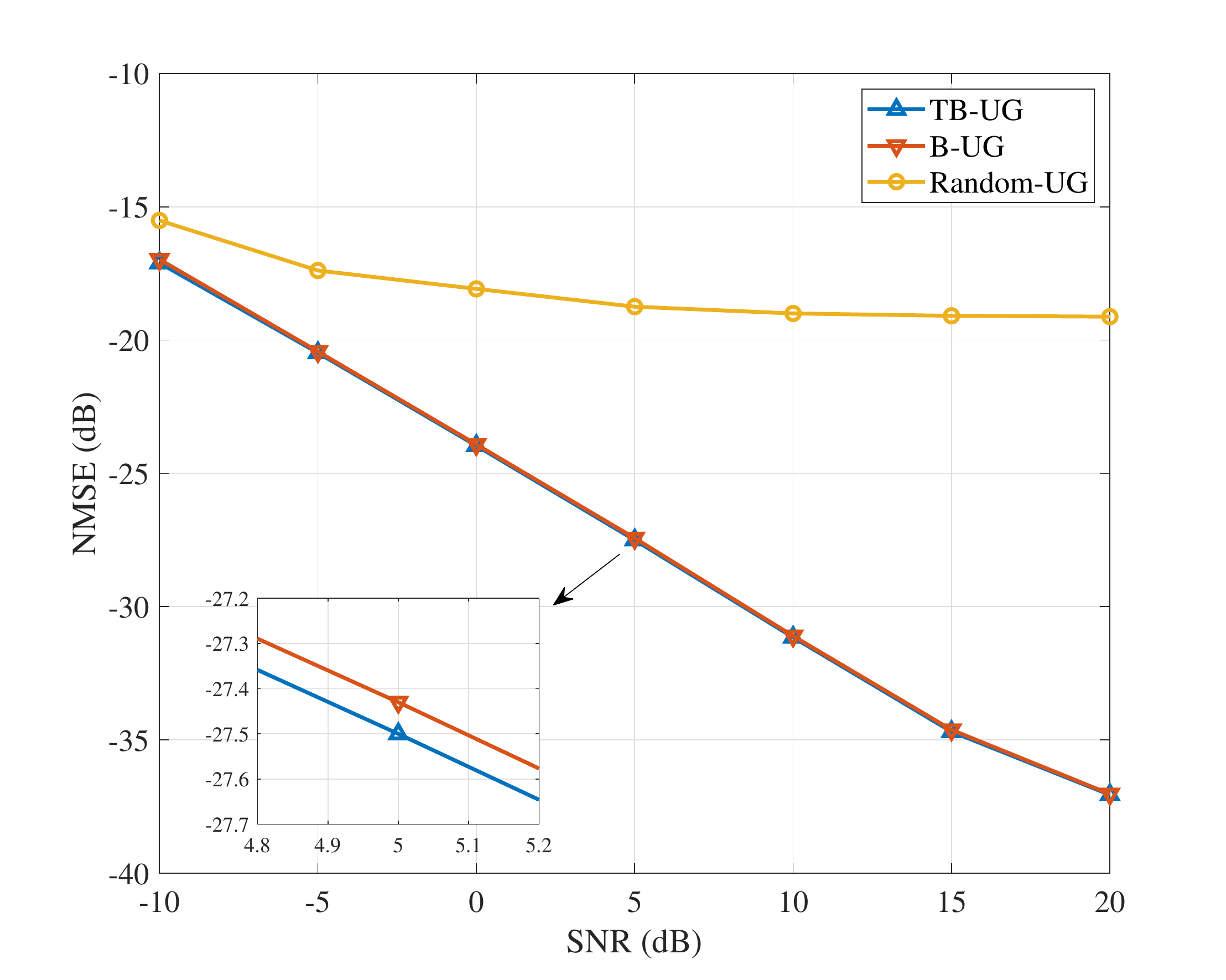}
			\label{fig-UG_all}
		\end{minipage}
		
	}
	\subfigure[]{
		\begin{minipage}{8.89cm}
			\centering
			\includegraphics[width=0.98\linewidth]{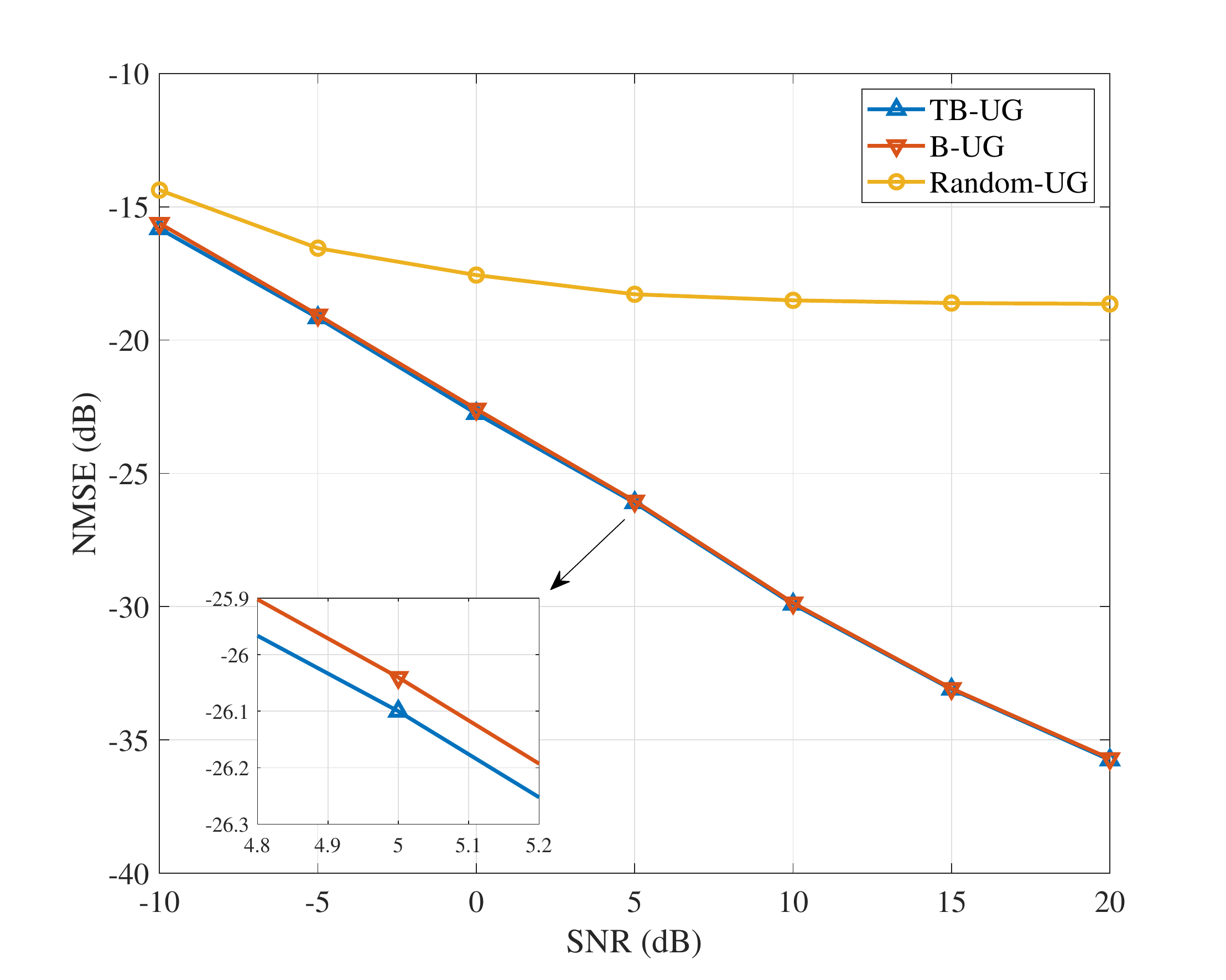}
			\label{fig-UG_last}
		\end{minipage}
	}
	\caption{The NMSE performance of different UT grouping algorithms. (a) NMSE of all timeslots; (b) NMSE of current timeslot.}
	\label{fig-nmse_ug}
\end{figure}

Fig. \ref{fig-nmse_ug} compares the NMSE performance of different UT grouping algorithms, where the UT velocity is 100 km/h, ${F_{{\rm{an}}}} = {F_{{\rm{de}}}} = {F_{{\rm{do}}}} = 2$, and the channel estimation algorithm is CBFEM-CE. From the figure, the proposed UT grouping algorithms are crucial to the channel estimation and have a significant performance gain over random-UG, especially when the SNR is high. This is because  there is severe inter-UT interference when the number of UTs is large, which can be effectively tackled by the proposed UT grouping algorithms.  Moreover, the performance of TB-UG is almost the same as that of B-UG while the computational complexity of ${\tilde \rho _{u,u'}}$ used in B-UG is much lower than that of ${\rho _{u,u'}}$ used in TB-UG.

\begin{figure}[htbp]
	\centering
	\includegraphics[width=0.98\linewidth]{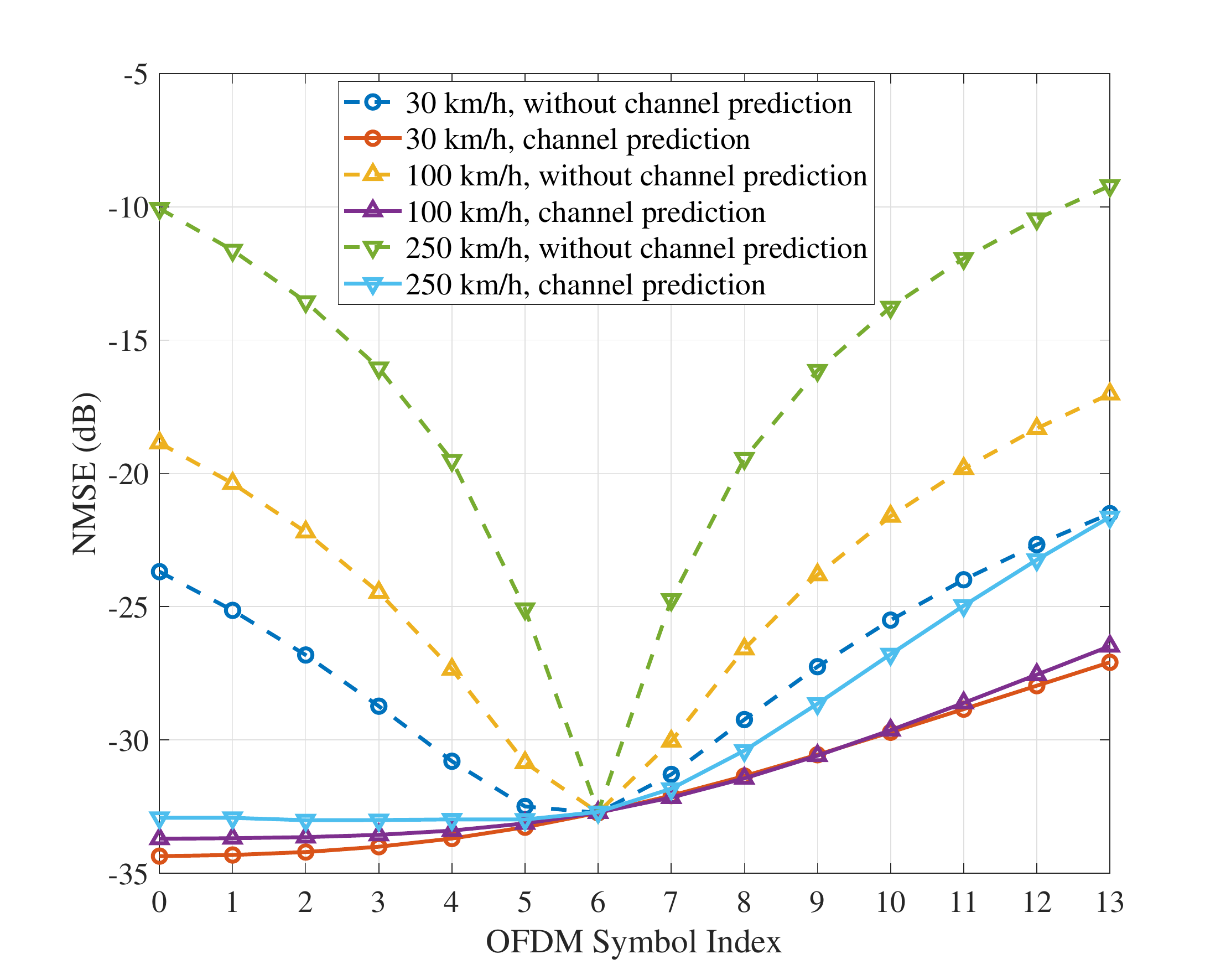}
	\caption{The NMSE performance versus OFDM symbol index under different UT velocities.}
	\label{fig-pre}
\end{figure}

Fig. 6 shows the NMSE performance of the proposed channel prediction method under different UT velocities, where the frame structure is as in Fig. 1, SNR $=$ 15 dB, the UT grouping algorithm is TB-UG, ${F_{{\rm{an}}}} = {F_{{\rm{de}}}} = {F_{{\rm{do}}}} = 2$, and the channel estimation algorithm is CBFEM-CE. The NMSE here refers to the NMSE between the channel obtained by channel prediction and the real channel of each OFDM symbol in the current timeslot.
We observe that when we directly apply the estimated channel at the pilot segment to the data segment, the NMSE performance will rapidly degrade as the delay between the pilot symbol and data symbols increases, especially when the UT velocity is fast. However, the proposed channel prediction method can acquire a more accurate CSI of the data segment due to the utilization of the estimated TB domain channel, which is conducive to DL transmit design and UL signal detection.

\section{Conclusions}
In this paper, we have investigated the channel acquisition for HF skywave massive MIMO-OFDM communications. 
We first established a TB based channel model using sampled triple steering vectors, each of which corresponds to a physical TB in the SFT domain. 
Then, based on the channel model, we investigated the optimal channel estimation for pilot segments and revealed the conditions for minimizing the NMSE of the channel estimate.
We showed that UTs with overlapping TB domain channels should be allocated pilot sequences with different phase shift factors while UTs with non-overlapping TB domain channels can reuse the same pilot sequence. Moreover, the pilot design was given, including UT grouping and pilot scheduling, and the channel prediction method for the data segment was provided based on the TB domain channel estimate.
To reduce the complexity of the channel estimation, we developed the CBFEM based channel estimation algorithm and its low-complexity implementation using CZT based on the structure of the TB matrix.
Simulation results verified the validity of the proposed CSI acquisition approach.

\appendices

\section{Proof of Theorem 1}
We start by presenting a lemma that is required in the subsequent proof.

\noindent {\bf Lemma 1}:  When one of the conditions in Theorem 1 satisfies, we have
\begin{align}\label{equ-appenB-1}
\mathop {\lim }\limits_{M,{N_{\rm{v}}},{N_{\rm{F}}} \to \infty } \frac{1}{{M{N_{\rm{v}}}{N_{\rm{F}}}}}{\bf{R}}_u^{{\rm{SFT,p}}}{{\bf{X}}_u^{\rm{H}}}{\bf{X}}_{u'}{\bf{R}}_{u'}^{{\rm{SFT,p}}} = {\bf{0}}.
\end{align}

{\itshape Proof}: According to (\ref{equ-Rsft}) and the fact that ${{\bf{X}}_u}{\bf{\tilde P}} \!=\! {\sigma _{\rm{p}}}{{\bf{X}}_{\rm{c}}}{\bf{\bar P}}{{\bf{S}}_u}$ we have
\begin{align}\label{equ-Lemma1-1}
&\mathop {\lim }\limits_{M,{N_{\rm{v}}},{N_{\rm{F}}} \to \infty }\! \frac{1}{{M{N_{\rm{v}}}{N_{\rm{F}}}}}{\bf{R}}_u^{{\rm{SFT,p}}}{\bf{X}}_u^{\rm{H}}{{\bf{X}}_{u'}}{\bf{R}}_{u'}^{{\rm{SFT,p}}} \nonumber \\
&=\!\!\!\! \mathop {\lim }\limits_{M,{N_{\rm{v}}},{N_{\rm{F}}} \to \infty }\! \frac{\sigma _{\rm{p}}^2}{{M{N_{\rm{v}}}{N_{\rm{F}}}}}{\bf{\tilde P}}\!\underbrace {{\bf{R}}_u^{{\rm{TB}}}{\bf{S}}_u^{\rm{H}}{{{\bf{\bar P}}}^{\rm{H}}}{\bf{X}}_{\rm{c}}^{\rm{H}}{{\bf{X}}_{\rm{c}}}{\bf{\bar P}}{{\bf{S}}_{u'}}{\bf{R}}_{u'}^{{\rm{TB}}}}_{\bf{\Phi }}\!{{\bf{\tilde P}}^{\rm{H}}}.
\end{align}
Then we have
\begin{align}
&\mathop {\lim }\limits_{M,{N_{\rm{v}}},{N_{\rm{F}}} \to \infty } {\frac{\sigma _{\rm{p}}^2}{{M{N_{\rm{v}}}{N_{\rm{F}}}}}\left[ {\bf{\Phi }} \right]_{a,b}} \nonumber \\
&= \mathop {\lim }\limits_{M,{N_{\rm{v}}},{N_{\rm{F}}} \to \infty } \sigma _{\rm{p}}^2{\left[ {{\bf{R}}_u^{{\rm{TB}}}} \right]_{a,a}}{\left[ {{\bf{R}}_{u'}^{{\rm{TB}}}} \right]_{b,b}}\nonumber\\
&\quad\quad\times\underbrace {\frac{1}{{{N_{\rm{F}}}}}\sum\limits_{{n_{\rm{F}}} = 0}^{{N_{\rm{F}}} - 1} {{e^{ - {\bar \jmath}2\pi \left({n_{\rm{F}}}{N_{\rm{S}}}+{n_{\rm{p}}}\right)\frac{{{N_{\rm{d}}}\left( {{n_{{\rm{do}}}} - {n'_{{\rm{do}}}}} \right)}}{{{N_{{\rm{do}}}}N}}}}} }_{{{\alpha _{\rm{T}}}\left( {{n_{{\rm{do}}}},{n'_{{\rm{do}}}}} \right)}} \nonumber \\
& \quad\quad\times\underbrace {\frac{1}{{{N_{\rm{v}}}}}\sum\limits_{k = {k_0}}^{{k_{{N_{\rm{v}}}}} - 1} {{e^{{\bar \jmath}2\pi k\frac{{{N_{\tau}}\left( {{n_{{\rm{de}}}} - {n'_{{\rm{de}}}} + {\phi _u} - {\phi _{u'}}} \right)}}{{{N_{{\rm{de}}}}{N_{\rm{v}}}}}}}} }_{{{\alpha _{\rm{F}}}\left( {{n_{{\rm{de}}}},{n'_{{\rm{de}}}},{\phi _u},{\phi _{u'}}} \right)}} \nonumber \\
&\quad\quad\times\underbrace {\frac{1}{M}\sum\limits_{m = 0}^{M - 1} {{e^{{\bar \jmath}2\pi \left( {{f_{\rm{c}}} + k\Delta f} \right)m\Delta \tau \frac{{2\left( {{n_{{\rm{an}}}} - {n'_{{\rm{an}}}}} \right)}}{{{N_{{\rm{an}}}}}}}}} }_{{{\alpha _{\rm{S}}}\left( {{n_{{\rm{an}}}},{n'_{{\rm{an}}}}},k \right)}},
\end{align}
where
\begin{align*}
{n_{{\rm{do}}}} \!=\! \left\lfloor {{a \mathord{\left/
			{\vphantom {a {\left( {{N_{{\rm{de}}}}{N_{{\rm{an}}}}} \right)}}} \right.
			\kern-\nulldelimiterspace} {\left( {{N_{{\rm{de}}}}{N_{{\rm{an}}}}} \right)}}} \right\rfloor,  \ &{n_{{\rm{de}}}} \!=\! {\left\langle {\left\lfloor {{a \mathord{\left/
					{\vphantom {a {{N_{{\rm{an}}}}}}} \right.
					\kern-\nulldelimiterspace} {{N_{{\rm{an}}}}}}} \right\rfloor } \right\rangle _{{N_{{\rm{de}}}}}}, \  {n_{{\rm{an}}}} \!=\! {\left\langle a \right\rangle _{{N_{{\rm{an}}}}}},\\
{n'_{{\rm{do}}}} \!= \!\left\lfloor {{b \mathord{\left/
			{\vphantom {b {\left( {{N_{{\rm{de}}}}{N_{{\rm{an}}}}} \right)}}} \right.
			\kern-\nulldelimiterspace} {\left( {{N_{{\rm{de}}}}{N_{{\rm{an}}}}} \right)}}} \right\rfloor, \  &{n'_{{\rm{de}}}} \!=\! {\left\langle {\left\lfloor {{b \mathord{\left/
					{\vphantom {b {{N_{{\rm{an}}}}}}} \right.
					\kern-\nulldelimiterspace} {{N_{{\rm{an}}}}}}} \right\rfloor } \right\rangle _{{N_{{\rm{de}}}}}}, \  {n'_{{\rm{an}}}} \!= \!{\left\langle b \right\rangle _{{N_{{\rm{an}}}}}},
\end{align*}
and ${\left\langle  \cdot  \right\rangle _N}$ denotes the modulo-$N$ operation.

When ${\bf{R}}_u^{{\rm{TB}}}{\bf{R}}_{u'}^{{\rm{TB}}} \ne {\bf{0}}$, there is an overlap between TB domain channels of UTs $u$ and $u'$, thus we let ${\phi _u} \ne {\phi _{u'}}$ so that ${n_{{\rm{de}}}} - {n'_{{\rm{de}}}} + {\phi _u} - {\phi _{u'}} \ne 0$ and ${{\alpha _{\rm{F}}}\left( {{n_{{\rm{de}}}},{n'_{{\rm{de}}}},{\phi _u},{\phi _{u'}}} \right)} = 0$ when ${N_{\rm{v}}} \to \infty $.
When ${\bf{R}}_u^{{\rm{TB}}}{\bf{R}}_{u'}^{{\rm{TB}}} = {\bf{0}}$, there is no overlap between TB domain channels of UTs $u$ and $u'$, which means that when both  ${\left[ {{\bf{R}}_u^{{\rm{TB}}}} \right]_{a,a}}$ and ${\left[ {{\bf{R}}_{u'}^{{\rm{TB}}}} \right]_{b,b}}$ are not equal to 0, we have $a \ne b$, thus at least one of ${n_{{\rm{do}}}} - {n'_{{\rm{do}}}} \ne 0$, ${n_{{\rm{de}}}} - {n'_{{\rm{de}}}} + {\phi _u} - {\phi _{u'}} \ne 0$, and ${n_{{\rm{an}}}} - {n'_{{\rm{an}}}}  \ne  0$ satisfies. In this case, at least one of ${{\alpha _{\rm{T}}}\left( {{n_{{\rm{do}}}},{n'_{{\rm{do}}}}} \right)}$, ${{\alpha _{\rm{F}}}\left( {{n_{{\rm{de}}}},{n'_{{\rm{de}}}},{\phi _u},{\phi _{u'}}} \right)}$ and ${{\alpha _{\rm{S}}}\left( {{n_{{\rm{an}}}},{n'_{{\rm{an}}}}},k \right)}$ is equal to 0 when $M,{N_{\rm{v}}},N \to \infty $.  In summary, when one of the conditions in Theorem 1 satisfies, 
\begin{equation}\label{equ-Lemma1-2}
\mathop {\lim }\limits_{M,{N_{\rm{v}}},N \to \infty }\frac{\sigma _{\rm{p}}^2}{{M{N_{\rm{v}}}{N_{\rm{F}}}}} { {\bf{\Phi}} } = \bf{0}.
\end{equation}
Substituting (\ref{equ-Lemma1-2}) into (\ref{equ-Lemma1-1}), we  obtain (\ref{equ-appenB-1}). This completes the proof of Lemma 1.

Next, we define that
\begin{equation}
{{\bf{\tilde C}}_u} \buildrel \Delta \over = {{\bf{X}}_u}{\bf{R}}_u^{{\rm{SFT,p}}}{\bf{X}}_u^{\rm{H}} + \sigma _{\rm{z}}^2{{\bf{I}}_{M{N_{\rm{v}}}{N_{\rm{F}}}}}.
\end{equation}
Due to the fact that ${\bf{C}} \succ {\bf{0}}$, ${{\bf{\tilde C}}_u} \succ {\bf{0}}$, and ${\bf{C}} - {{\bf{\tilde C}}_u} \succeq {\bf{0}}$, we can obtain that ${{{\bf{\tilde C}}_u}^{ - 1}} \!-\! {{\bf{C}}^{ - 1}} \succeq {\bf{0}}$ \cite{seber2008matrix}. Therefore, ${\bf{R}}_u^{{\rm{SFT,p}}}{\bf{X}}_u^{\rm{H}}\left( {{{{\bf{\tilde C}}_u}^{ - 1}} - {{\bf{C}}^{ - 1}}} \right){{\bf{X}}_u}{\bf{R}}_u^{{\rm{SFT,p}}} \succeq {\bf{0}}$, and
\begin{align}\label{equ-appen1}
&{\rm{NMSE}}  \sum\limits_{u = 0}^{U - 1} \frac{1}{{M{N_{\rm{v}}}{N_{\rm{F}}}U{\vartheta _u}}}{\rm{tr}}\Big\{ {\bf{R}}_u^{{\rm{SFT,p}}} \nonumber \\
&\qquad\qquad\qquad\qquad\quad-{\bf{R}}_u^{{\rm{SFT,p}}}{\bf{X}}_u^{\rm{H}}{{\bf{\tilde C}}_u^{ - 1}}{{\bf{X}}_u}{\bf{R}}_u^{{\rm{SFT,p}}} \Big\}.
\end{align}
According to Lemma 1, when one of the conditions in Theorem 1 satisfies, we have
\begin{align}\label{equ-appenB2}
&\mathop {\lim }\limits_{M,{N_{\rm{v}}},{N_{\rm{F}}} \to \infty } \frac{1}{{M{N_{\rm{v}}}{N_{\rm{F}}}}}{\bf{R}}_u^{{\rm{SFT}},{\rm{p}}}{\bf{X}}_u^{\rm{H}}{\bf{C}}{{\bf{X}}_u}{\bf{R}}_u^{{\rm{SFT}},{\rm{p}}} \nonumber \\
& \qquad= \mathop {\lim }\limits_{M,{N_{\rm{v}}},{N_{\rm{F}}} \to \infty } \frac{1}{{M{N_{\rm{v}}}{N_{\rm{F}}}}}{\bf{R}}_u^{{\rm{SFT}},{\rm{p}}}{\bf{X}}_u^{\rm{H}}{{{\bf{\tilde C}}}_u}{{\bf{X}}_u}{\bf{R}}_u^{{\rm{SFT}},{\rm{p}}}\nonumber\\
&\Rightarrow \!\!\!\!\!\!\!\!\mathop {\lim }\limits_{M,{N_{\rm{v}}},{N_{\rm{F}}} \to \infty } \!\!\frac{1}{{M\!{N_{\rm{v}}}\!{N_{\rm{F}}}}}{\rm{tr}}\!\left\{\! {{\bf{R}}_u^{{\rm{SFT}}\!,{\rm{p}}}\! -\! {\bf{R}}_u^{{\rm{SFT}}\!,{\rm{p}}}{\bf{X}}_u^{\rm{H}}{{\bf{C}}^{ -\! 1}}{{\bf{X}}_u}{\bf{R}}_u^{{\rm{SFT}},{\rm{p}}}}\! \right\} \nonumber\\
&\ = \!\!\!\!\!\!\!\!\!\mathop {\lim }\limits_{M,{N_{\rm{v}}},{N_{\rm{F}}} \to \infty } \!\!\frac{1}{{M\!{N_{\rm{v}}}\!{N_{\rm{F}}}}}\!{\rm{tr}}\Big\{\! {{\bf{R}}_u^{{\rm{SFT}},{\rm{p}}}\! \!-\! {\bf{R}}_u^{{\rm{SFT}}\!,{\rm{p}}}{\bf{X}}_u^{\rm{H}}{\bf{\tilde C}}_u^{ - \!1}{{\bf{X}}_u}{\bf{R}}_u^{{\rm{SFT}}\!,{\rm{p}}}} \!\Big\}.
\end{align}
Substituting (\ref{equ-appenB2}) into (\ref{equ-nmse-ini}), the equality in (\ref{equ-appen1}) holds and the NMSE is reduced to the minimum value in (\ref{equ-nmse_min}). This completes the proof.

\section{Proof of Theorem 2}
We start by presenting a lemma that is required in the subsequent proof.

\noindent {\bf Lemma 2}:  For arbitrary ${N_{\rm{v}}}$ and ${N_{\rm{F}}}$,  when $u' \notin {\mathcal{I}_u}$, we have
\begin{align}\label{equ-appenC-1}
\mathop {\lim }\limits_{M \to \infty } \frac{1}{{M{N_{\rm{v}}}{N_{\rm{F}}}}}{\bf{R}}_u^{{\rm{SFT,p}}}{{\bf{X}}_u^{\rm{H}}}{\bf{X}}_{u'}{\bf{R}}_{u'}^{{\rm{SFT,p}}} = {\bf{0}}.
\end{align}

{\itshape Proof}: When ${N_{\rm{v}}}$ and ${N_{\rm{F}}}$ are arbitrary and  only ${M \to \infty }$, we can obtain that
\begin{equation}\label{equ-Lemma2-1}
\mathop {\lim }\limits_{M \to \infty }\! \frac{1}{{M\!{N_{\rm{v}}}\!{N_{\rm{F}}}}}{\bf{R}}_u^{{\rm{SFT\!,p}}}{{\bf{X}}_u^{\rm{H}}}{\bf{X}}_{u'}{\bf{R}}_{u'}^{{\rm{SFT\!,p}}} \!=\!\!\! \mathop {\lim }\limits_{M \to \infty } \!\!\frac{\sigma _{\rm{p}}^2}{{M\!{N_{\rm{v}}}\!{N_{\rm{F}}}}}{\bf{\tilde P \tilde \Phi }}{{\bf{\tilde P}}^{\rm{H}}},
\end{equation}
where
\begin{align}
&\mathop {\lim }\limits_{M \to \infty } \frac{\sigma _{\rm{p}}^2}{{M{N_{\rm{v}}}{N_{\rm{F}}}}}{\left[ {\bf{\tilde \Phi }} \right]_{a,b}} \nonumber \\
&= \!\!\mathop {\lim }\limits_{M \to \infty }\!\!\sigma _{\rm{p}}^2 \!\!\!\!\sum\limits_{i = 0}^{{N_{{\rm{de}}}}{N_{{\rm{do}}}} - 1} \!{\sum\limits_{j = 0}^{{N_{{\rm{de}}}}{N_{{\rm{do}}}} - 1}\!\!\!\!\!\!\!\! {{{\left[ {{\bf{R}}_u^{{\rm{TB}}}} \right]}_{a,i{N_{{\rm{an}}}} + {n_{{\rm{an}}}}}}\!{{\left[ {{\bf{R}}_{u'}^{{\rm{TB}}}} \right]}_{j{N_{{\rm{an}}}} + {n'_{{\rm{an}}}},b}}} } \nonumber\\
&\ \times\! {\alpha _{\rm{T}}}\!\left( {{n''_{{\rm{do}}}},{n'''_{{\rm{do}}}}} \right){\alpha _{\rm{F}}}\!\left( {{n''_{{\rm{de}}}},{n'''_{{\rm{de}}}},{\phi _u},{\phi _{u'}}} \!\right){\alpha _{\rm{S}}}\!\left( {{n_{{\rm{an}}}},{n'_{{\rm{an}}}}},k \right),
\end{align}
where 
\begin{align*}
{n''_{{\rm{do}}}}\! = \!\left\lfloor {{i \mathord{\left/
			{\vphantom {i {{N_{{\rm{de}}}}}}} \right.
			\kern-\nulldelimiterspace} {{N_{{\rm{de}}}}}}} \right\rfloor,\  {n''_{{\rm{de}}}} \!= \!{\left\langle i \right\rangle _{{N_{{\rm{de}}}}}},\ {n'''_{{\rm{do}}}}\! =\! \left\lfloor {{j \mathord{\left/
			{\vphantom {j {{N_{{\rm{de}}}}}}} \right.
			\kern-\nulldelimiterspace} {{N_{{\rm{de}}}}}}} \right\rfloor,\  {n'''_{{\rm{de}}}} \!=\! {\left\langle j \right\rangle _{{N_{{\rm{de}}}}}},
\end{align*}
and ${n_{{\rm{an}}}}$, ${n'_{{\rm{an}}}}$, ${{\alpha _{\rm{T}}}\left( {{n''_{{\rm{do}}}},{n'''_{{\rm{do}}}}} \right)}$, ${{\alpha _{\rm{F}}}\left( {{n''_{{\rm{de}}}},{n'''_{{\rm{de}}}},{\phi _u},{\phi _{u'}}} \right)}$ and ${{\alpha _{\rm{S}}}\left( {{n_{{\rm{an}}}},{n'_{{\rm{an}}}}},k \right)}$ are similar to the definitions in Appendix A. 

When $u' \notin {\mathcal{I}_u}$, the TB domain channel of UTs $u$ and $u'$ is non-overlapping along the spatial-beam dimension, which means that when both  ${{{\left[ {{\bf{R}}_u^{{\rm{TB}}}} \right]}_{a,i{N_{{\rm{an}}}} + {n_{{\rm{an}}}}}}}$ and ${{{\left[ {{\bf{R}}_{u'}^{{\rm{TB}}}} \right]}_{j{N_{{\rm{an}}}} + {n'_{{\rm{an}}}},b}}}$ are not equal to 0, we have ${n_{{\rm{an}}}} \ne {n'_{{\rm{an}}}}$, thus ${\alpha _{\rm{S}}}\left( {{n_{{\rm{an}}}},{n'_{{\rm{an}}}}},k \right) = 0$ when $M \to \infty $. Therefore, we have
\begin{equation}\label{equ-Lemma2-2}
\mathop {\lim }\limits_{M \to \infty } \frac{\sigma _{\rm{p}}^2}{{M{N_{\rm{v}}}{N_{\rm{F}}}}}{{\bf{\tilde\Phi}} } = \bf{0}.
\end{equation}
Substituting (\ref{equ-Lemma2-2}) into (\ref{equ-Lemma2-1}), we  obtain (\ref{equ-appenC-1}). This completes the proof of Lemma 2.

Then we have
\begin{align}\label{equ-appen2}
&\mathop {\lim }\limits_{M \to \infty } \frac{1}{{M{N_{\rm{v}}}{N_{\rm{F}}}}}{\bf{R}}_u^{{\rm{SFT,p}}}{\bf{X}}_u^{\rm{H}}{\bf{C}}{{\bf{X}}_u}{\bf{R}}_u^{{\rm{SFT,p}}}\nonumber \\
&\mathop  = \limits^{\left( {\rm{a}} \right)} \mathop {\lim }\limits_{M \to \infty } \frac{1}{{M{N_{\rm{v}}}{N_{\rm{F}}}}}{\bf{R}}_u^{{\rm{SFT,p}}}{\bf{X}}_u^{\rm{H}}{\bf{\bar C}}_u{{\bf{X}}_u}{\bf{R}}_u^{{\rm{SFT,p}}}\nonumber\\
\Rightarrow &\mathop {\lim }\limits_{M \to \infty }\! \frac{1}{{M{N_{\rm{v}}}{N_{\rm{F}}}}}{\rm{tr}}\!\left\{\! {{\bf{R}}_u^{{\rm{SFT,p}}} \!-\! {\bf{R}}_u^{{\rm{SFT,p}}}{\bf{X}}_u^{\rm{H}}{{\bf{C}}^{ - 1}}{{\bf{X}}_u}{\bf{R}}_u^{{\rm{SFT,p}}}} \right\} \nonumber\\
&\!\!\!= \mathop {\lim }\limits_{M \to \infty }\! \frac{1}{{M{N_{\rm{v}}}{N_{\rm{F}}}}}{\rm{tr}}\!\left\{\! {{\bf{R}}_u^{{\rm{SFT\!,p}}} \!\!-\! {\bf{R}}_u^{{\rm{SFT\!,p}}}{\bf{X}}_u^{\rm{H}}{{{\bf{\bar C}}}_u^{ -\! 1}}{{\bf{X}}_u}{\bf{R}}_u^{{\rm{SFT\!,p}}}} \right\},
\end{align}
where (a) follows from Lemma 2. Substituting (\ref{equ-appen2}) into (\ref{equ-nmse-ini}), we can obtain (\ref{equ-theorem2}) . This concludes the proof.

\section{Derivation of CBFEM based channel estimation algorithm}
The Lagrange function of the (\ref{equ-Bethe5}) can be given by
\begin{align}\label{equ-C-lag}
&{L_{\rm{B}}} = {F_{\rm{B}}}\nonumber \\
&  + \!\!\!\!\!\sum\limits_{i = 0}^{M{N_{\rm{v}}}{N_{\rm{F}}} - 1} \!\!\!\!\!\!{2{\mathop{\rm Re}\nolimits} \left\{ {{{\left( {\tau _i^{w,{b_y}}} \right)}^ * }\left( {{\mathbb{E}}\left\{ {{w_i}\left| {{q_{w,i}}} \right.} \right\} - {\mathbb{E}}\left\{ {{w_i}\left| {{b_{y,i}}} \right.} \right\}} \right)} \right\}} \nonumber\\
&+\!\!\!\!\! \sum\limits_{i = 0}^{M{N_{\rm{v}}}{N_{\rm{F}}} - 1} \!\!\!\!\!\!{2{\mathop{\rm Re}\nolimits} \left\{ {{{\left( {\tau _i^{w,{b_w}}} \right)}^ * }\left( {{\mathbb{E}}\left\{ {{w_i}\left| {{q_{w,i}}} \right.} \right\} - {\mathbb{E}}\left\{ {{w_i}\left| {{b_{w,i}}} \right.} \right\}} \right)} \right\}} \nonumber\\
&+ \!\!\!\!\!\!\!\!\!\sum\limits_{i = 0}^{M{N_{\rm{v}}}{N_{\rm{F}}} - 1} \!{\sum\limits_{j = 0}^{{N_{\rm{an}}}{N_{\rm{de}}}{N_{\rm{do}}}U - 1} \!\!\!\!\!\!\!\!\!\!\!\!\!\!{2{\mathop{\rm Re}\nolimits} \!\left\{\! \!{{{\left(\! {\tau _{i,j}^{h,{b_w}}} \!\right)}^{\!\!*} }\!\!\!\left( {{\mathbb{E}}\!\left\{\! {h_j^{{\rm{TB}}}\!\left| {{q_{h,j}}} \!\right.} \!\right\}\!\! -\! {\mathbb{E}}\!\left\{\! {h_j^{{\rm{TB}}}\!\left| {{b_{w,i}}} \right.} \right\}} \!\right)} \!\!\right\}} } \nonumber\\
&+\!\!\!\!\!\!\! \sum\limits_{j = 0}^{{N_{\rm{an}}}{N_{\rm{de}}}{N_{\rm{do}}}U - 1}\!\!\!\!\!\!\!\!\!\!\!\!{2{\mathop{\rm Re}\nolimits} \left\{\! {{{\left(\! {\tau _j^{h,{b_h}}} \!\right)}^ * }\left( {{\mathbb{E}}\left\{ {h_j^{{\rm{TB}}}\!\left| {{q_{h,j}}} \!\right.} \right\} - {\mathbb{E}}\left\{ {h_j^{{\rm{TB}}}\!\left| {{b_{h,j}}} \right.} \right\}} \right)} \right\}} \nonumber\\
&+ \sum\limits_{i = 0}^{M{N_{\rm{v}}}{N_{\rm{F}}} - 1} {\eta _i^{w,{b_y}}\left( {{\mathbb{E}}\left\{ {{{\left| {{w_i}} \right|}^2}\left| {{q_{w,i}}} \right.} \right\} - {\mathbb{E}}\left\{ {{{\left| {{w_i}} \right|}^2}\left| {{b_{y,i}}} \right.} \right\}} \right)} \nonumber\\
&+ \sum\limits_{i = 0}^{M{N_{\rm{v}}}{N_{\rm{F}}} - 1} {\eta _i^{w,{b_w}}\left( {{\mathbb{E}}\left\{ {{{\left| {{w_i}} \right|}^2}\left| {{q_{w,i}}} \right.} \right\} - {\mathbb{E}}\left\{ {{{\left| {{w_i}} \right|}^2}\left| {{b_{w,i}}} \right.} \right\}} \right)} \nonumber\\
&+\!\!\!\!\!\!\!\! \sum\limits_{j = 0}^{{N_{\rm{an}}}{N_{\rm{de}}}{N_{\rm{do}}}U \!-\! 1}\!\!\!\!\!\!\!\!\!\!\!\!\!\! {\eta _j^{h,{b_w}}\!\!\left(\!\! {M{N_{\rm{v}}}{N_{\rm{F}}}{\mathbb{E}}\!\left\{\! {{{\left| {h_j^{{\rm{TB}}}} \right|}^2}\!\left| {{q_{h,j}}} \!\right.} \!\right\} \!-\!\!\!\!\!\!\!\! \sum\limits_{i = 0}^{M{N_{\rm{v}}}{N_{\rm{F}}}\! -\! 1} \!\!\!\!\!\!\!\!{{\mathbb{E}}\!\left\{\! {{{\left| {h_j^{{\rm{TB}}}} \right|}^2}\!\left| {{b_{w,i}}} \right.} \!\right\}} } \!\!\right)} \nonumber\\
&+\!\!\!\!\!\! \sum\limits_{j = 0}^{{N_{\rm{an}}}{N_{\rm{de}}}{N_{\rm{do}}}U - 1}\!\!\!\!\!\!\!\!\! {\eta _j^{h,{b_h}}\left( {{\mathbb{E}}\left\{ {{{\left| {h_j^{{\rm{TB}}}} \right|}^2}\left| {{q_{h,j}}} \right.} \right\} - {\mathbb{E}}\left\{ {{{\left| {h_j^{{\rm{TB}}}} \right|}^2}\left| {{b_{h,j}}} \right.} \right\}} \right)} .
\end{align}
Then by setting the first-order derivatives of (\ref{equ-C-lag}) for each belief equal to zeros, a series of fixed-point equations can be obtained  as follows.
\begin{equation}
{b_{y,i}} \propto p\left( {{y_i}\left| {{w_i}} \right.} \right)\mathcal{CN}\left( {{w_i}; - \frac{{\tau _i^{w,{b_y}}}}{{\eta _i^{w,{b_y}}}}, - \frac{1}{{\eta _i^{w,{b_y}}}}} \right),
\end{equation}
\begin{equation}
{q_{w,i}} \propto \mathcal{CN}\left( {{w_i}; - \frac{{\tau _i^{w,{b_y}} + \tau _i^{w,{b_w}}}}{{\eta _i^{w,{b_y}} + \eta _i^{w,{b_w}}}}, - \frac{1}{{\eta _i^{w,{b_y}} + \eta _i^{w,{b_w}}}}} \right),
\end{equation}
\begin{align}
{b_{w,i}}& \propto p\left( {{w_i}\left| {{{\bf{h}}^{{\rm{TB}}}}} \right.} \right)\mathcal{CN}\left( {{w_i}; - \frac{{\tau _i^{w,{b_w}}}}{{\eta _i^{w,{b_w}}}}, - \frac{1}{{\eta _i^{w,{b_w}}}}} \right) \nonumber \\
& \qquad\times\!\!\!\prod\limits_{j = 0}^{{N_{\rm{an}}}{N_{\rm{de}}}{N_{\rm{do}}}U - 1}\!\!\!\!\!\!\! {\mathcal{CN}\left( {h_j^{{\rm{TB}}}; - \frac{{\tau _{i,j}^{h,{b_w}}}}{{\eta _j^{h,{b_w}}}}, - \frac{1}{{\eta _j^{h,{b_w}}}}} \right)} ,
\end{align}
\begin{equation}
{q_{h,j}} \!\propto\! \mathcal{CN}\!\!\left(\!\! {h_j^{\!{\rm{TB}}}\!;\! - \frac{{\sum\limits_{i = 0}^{M\!{N_{\rm{v}}}\!{N_{\rm{F}}} \!-\! 1} \!\!\!\!{\tau _{i,j}^{h\!,{b_w}} \!+ } \tau _j^{h\!,{b_h}}}}{{M\!{N_{\rm{v}}}\!{N_{\rm{F}}}\eta _j^{h\!,{b_w}}\!\!\! +\! \eta _j^{h\!,{b_h}}}},\! - \frac{{M{N_{\rm{v}}}{N_{\rm{F}}}}}{{M\!\!{N_{\rm{v}}}\!{N_{\rm{F}}}\eta _j^{h\!,{b_w}} \!\!\!+\!\! \eta _j^{h\!,{b_h}}}}} \!\!\!\right),
\end{equation}
\begin{equation}\label{equ-appenC1}
{b_{h,j}} \propto p\left( {h_j^{{\rm{TB}}}} \right)\mathcal{CN}\left( {h_j^{{\rm{TB}}}; - \frac{{\tau _j^{h,{b_h}}}}{{\eta _j^{h,{b_h}}}}, - \frac{1}{{\eta _j^{h,{b_h}}}}} \right).
\end{equation}

Next, for constraints (\ref{equ-Bethe4}), let (XX-1), (XX-2), (XX-3) denote the equation between the first and the second term, the second and the third term, the first  and third term, respectively, for brevity. For example, (\ref{equ-Bethe4-1}-1) denotes ${\mathbb{E}}\left\{ {{w_i}\left| {{b_{y,i}}} \right.} \right\} = {\mathbb{E}}\left\{ {{w_i}\left| {{b_{w,i}}} \right.} \right\}$, (\ref{equ-Bethe4-1}-2) denotes ${\mathbb{E}}\left\{ {{w_i}\left| {{b_{w,i}}} \right.} \right\} = {\mathbb{E}}\left\{ {{w_i}\left| {{q_{w,i}}} \right.} \right\}$ and (\ref{equ-Bethe4-1}-3) denotes ${\mathbb{E}}\left\{ {{w_i}\left| {{b_{y,i}}} \right.} \right\} = {\mathbb{E}}\left\{ {{w_i}\left| {{q_{w,i}}} \right.} \right\}$. let $a_{ij}$ denote the $(i,j)$-th element of $\bf{A}$. Note that the modulus of of each element of ${\bf{A}}$ is the same, i.e., ${\sigma _{\rm{p}}}$.

According to (\ref{equ-Bethe4-2}-3) and (\ref{equ-Bethe4-4}-3), we can obtain that
\begin{align}\label{equ-C2}
 &\eta _j^{h,{b_w}} =  - \frac{1}{{{\rm{Var}}\left\{ {h_j^{{\rm{TB}}}\left| {{b_{h,j}}} \right.} \right\}}} - \frac{{\eta _j^{h,{b_h}}}}{{M{N_{\rm{v}}}{N_{\rm{F}}}}}.
\end{align}
According to (\ref{equ-Bethe4-1}-2) and (\ref{equ-Bethe4-3}-2), we can obtain that
\begin{align}\label{equ-C3}
\eta _i^{w,{b_y}} = {\left( {\sum\limits_{j = 0}^{{N_{\rm{an}}}{N_{\rm{de}}}{N_{\rm{do}}}U - 1} {\frac{{\sigma _{\rm{p}}^2} }{{\eta _j^{h,{b_w}}}}} } \right)^{ - 1}} \buildrel \Delta \over = {{\tilde \eta }^{w,{b_y}}} .
\end{align}
According to (\ref{equ-Bethe4-1}-3) , (\ref{equ-Bethe4-3}-3) and $p\left( {{y_i}\left| {{w_i}} \right.} \right) \propto \mathcal{CN}\left( {{y_i};{w_i},\sigma _{\rm{z}}^2} \right)$, we can obtain that $\eta _i^{w,{b_w}} =  - {1 \mathord{\left/
		{\vphantom {1 {\sigma _{\rm{z}}^2}}} \right.
		\kern-\nulldelimiterspace} {\sigma _{\rm{z}}^2}}$ and $\tau _i^{w,{b_w}} = {{{y_i}} \mathord{\left/
		{\vphantom {{{y_i}} {\sigma _{\rm{z}}^2}}} \right.
		\kern-\nulldelimiterspace} {\sigma _{\rm{z}}^2}}$. Then according to (\ref{equ-Bethe4-2}-2), (\ref{equ-Bethe4-4}-2) and (\ref{equ-C3}), we can obtain that
\begin{align}\label{equ-C4}
\eta _j^{h,{b_h}} = {\left( {\frac{{{{\left( {{{\tilde \eta }^{w,{b_y}}}} \right)}^{ - 1}} - \sigma _{\rm{z}}^2}}{{M{N_{\rm{v}}}{N_{\rm{F}}}}{\sigma _{\rm{p}}^2}} - \frac{1}{{M{N_{\rm{v}}}{N_{\rm{F}}}\eta _j^{h,{b_w}}}}} \right)^{ - 1}},
\end{align}
\begin{align}\label{equ-C7}
\tau _j^{h,{b_h}}\! =\!\!\!\!\! \sum\limits_{i = 0}^{M{N_{\rm{v}}}{N_{\rm{F}}} - 1}\!\!\!\! { - \frac{{a_{ij}^ * \left( {{y_i} + {\psi _i}} \right) - {\sigma _{\rm{p}}^2} \frac{{\tau _{i,j}^{h,{b_w}}}}{{\eta _j^{h,{b_w}}}}}}{{{{\left( {{\tilde \eta }^{w,{b_y}}} \right)}^{ - 1}} - \sigma _{\rm{z}}^2 - \frac{{\sigma _{\rm{p}}^2} }{{\eta _j^{h,{b_w}}}}}}} \! =\!\!\!\!\! \sum\limits_{i = 0}^{M{N_{\rm{v}}}{N_{\rm{F}}} - 1} \!\!\!\!{{\mu _{i,j}}} ,
\end{align}
where
\begin{equation} \label{equ-psi}
{\psi _i} \buildrel \Delta \over = \sum\limits_{j = 0}^{{N_{\rm{an}}}{N_{\rm{de}}}{N_{\rm{do}}}U - 1} {{a_{ij}}\frac{{\tau _{i,j}^{h,{b_w}}}}{{\eta _j^{h,{b_w}}}}} ,
\end{equation}
\begin{equation}
{\mu _{i,j}} \buildrel \Delta \over =  - \frac{{a_{ij}^ * \left( {{y_i} + {\psi _i}} \right) - {\sigma _{\rm{p}}^2} \frac{{\tau _{i,j}^{h,{b_w}}}}{{\eta _j^{h,{b_w}}}}}}{{{{\left( {{\tilde \eta }^{w,{b_y}}} \right)}^{ - 1}} - \sigma _{\rm{z}}^2 - \frac{{\sigma _{\rm{p}}^2} }{{\eta _j^{h,{b_w}}}}}}.
\end{equation} 
According to (\ref{equ-C4}) and (\ref{equ-C7}), we have 
\begin{equation}\label{equ-appenC2}
{\varpi  _j} \!\buildrel \Delta \over =\!  - \frac{{\tau _j^{h,{b_h}}}}{{\eta _j^{h,{b_h}}}} \!=\! \frac{1}{{M{N_{\rm{v}}}{N_{\rm{F}}}\sigma _{\rm{p}}^2}}\!\!\!\!\!\!\!\!\sum\limits_{i = 0}^{M{N_{\rm{v}}}{N_{\rm{F}}} - 1}\!\!\! \left(\!\!{a_{ij}^*\left( {{y_i} + {\psi _i}} \right) \!-\! \sigma _{\rm{p}}^2\frac{{\tau _{i,j}^{h,{b_w}}}}{{\eta _j^{h,{b_w}}}}}\!\!\right) .
\end{equation}
According to (\ref{equ-Bethe4-2}-3) and (\ref{equ-Bethe4-4}-3), we can obtain that
\begin{align}\label{equ-C5}
\sum\nolimits_{i = 0}^{M{N_{\rm{v}}}{N_{\rm{F}}} - 1} \!\!{\tau _{i,j}^{h,{b_w}}}& \! =\! \frac{{M{N_{\rm{v}}}{N_{\rm{F}}}{\mathbb{E}}\left\{ {h_j^{{\rm{TB}}}\left| {{b_{h,j}}} \right.} \right\}}}{{{\rm{Var}}\left\{ {h_j^{{\rm{TB}}}\left| {{q_{h,j}}} \right.} \right\}}} \!-\! \tau _j^{h,{b_h}}  \nonumber \\
&\!= \!\frac{{M{N_{\rm{v}}}{N_{\rm{F}}}{\mathbb{E}}\left\{ {h_j^{{\rm{TB}}}\left| {{b_{h,j}}} \right.} \right\}}}{{{\rm{Var}}\left\{ {h_j^{{\rm{TB}}}\left| {{b_{h,j}}} \right.} \right\}}}\! -\! \tau _j^{h,{b_h}}.
\end{align}
According to (\ref{equ-C7}), (\ref{equ-C5}) and the constraint $\mathbb{E}\left\{ {h_j^{{\rm{TB}}}\left| {{b_{w,i}}} \right.} \right\} = \mathbb{E}\left\{ {h_j^{{\rm{TB}}}\left| {{b_{w,i'}}} \right.} \right\}$,  we can obtain that
\begin{align}\label{equ-C8}
\tau _{i,j}^{h,{b_w}} = \frac{{{\mathbb{E}}\left\{ {h_j^{{\rm{TB}}}\left| {{b_{h,j}}} \right.} \right\}}}{{{\rm{Var}}\left\{ {h_j^{{\rm{TB}}}\left| {{b_{h,j}}} \right.} \right\}}} - \tau _j^{h,{b_h}} + \sum\limits_{i' = 0,i' \ne i}^{M{N_{\rm{v}}}{N_{\rm{F}}} - 1} {{\mu _{i',j}}} .
\end{align}
In HF skywave massive MIMO-OFDM communications, ${M{N_{\rm{v}}}{N_{\rm{F}}}}$ is usually very large. Hence we approximate that $\sum\limits_{i' = 0,i' \ne i}^{M{N_{\rm{v}}}{N_{\rm{F}}} - 1} {{\mu _{i',j}}}  \approx \sum\limits_{i = 0}^{M{N_{\rm{v}}}{N_{\rm{F}}} - 1} {{\mu _{i,j}}} $. The approximation here is conducive to the derivation of the  low-complexity implementation, and the resulting channel estimation algorithm still has a satisfactory performance, as shown in simulation results.
Then $\tau _{i,j}^{h,{b_w}}$ is rewritten as
\begin{equation}\label{equ-C9}
\tau _{i,j}^{h,{b_w}} = \frac{{{\mathbb{E}}\left\{ {h_j^{{\rm{TB}}}\left| {{b_{h,j}}} \right.} \right\}}}{{{\rm{Var}}\left\{ {h_j^{{\rm{TB}}}\left| {{b_{h,j}}} \right.} \right\}}} \buildrel \Delta \over = \tilde \tau _j^{h,{b_w}}.
\end{equation}
We define ${{\bm{\eta }}^{h,{b_w}}}$, ${{\bm{\eta }}^{h,{b_h}}}$, ${{\bm{\tilde \tau }}^{h,{b_w}}}$ and ${\bm{\varpi }}$, whose $j$-th element are ${{{\eta }}_j^{h,{b_w}}}$, ${{{\eta }}_j^{h,{b_h}}}$, ${{{\tilde \tau }}_j^{h,{b_w}}}$ and ${\varpi _j}$, respectively. In addition, we define ${\bm{\psi }}$, whose $i$-th element is ${{\psi }}_i$. Then (\ref{equ-appenC1}), (\ref{equ-C2}), (\ref{equ-C3}), (\ref{equ-C4}), (\ref{equ-psi}), (\ref{equ-appenC2}) and (\ref{equ-C9}) can be rewritten in vector and matrix form and constitute the CBFEM based channel estimation algorithm.

\bibliographystyle{IEEEtran}
\bibliography{reference}

\end{document}